\documentclass[11pt,a4paper]{article}
\pdfoutput=1

\usepackage[colorlinks=true, linkcolor=black!50!blue, urlcolor=blue, citecolor=blue, anchorcolor=blue]{hyperref}
\usepackage[font=small,labelfont=bf,margin=0mm,labelsep=period,tableposition=top]{caption}
\usepackage[a4paper,top=1.5cm,bottom=2cm,left=1.5cm,right=1.5cm,bindingoffset=0mm]{geometry}

\usepackage{placeins,cite}
\usepackage{graphicx}
\usepackage{float}
\usepackage{afterpage}
\usepackage{epsfig}
\usepackage{amssymb}
\usepackage{amsmath}
\usepackage{bm}
\usepackage{multirow}
\usepackage{url}
\usepackage{xcolor}
\usepackage{ulem}
\usepackage{url}
\usepackage{booktabs,multirow}
\usepackage{textcomp}
\usepackage{enumitem}
\graphicspath{{../}{./figures/}}
\usepackage{placeins}

\makeatletter
\def\thickhline{%
             \noalign{\ifnum0 =`}\fi\hrule \@height \thickarrayrulewidth \futurelet
             \reserved@a\@xthickhline}
\def\@xthickhline{\ifx\reserved@a\thickhline
                \vskip\doublerulesep
                \vskip -\thickarrayrulewidth
                \fi
                \ifnum0 =`{\fi}}
\makeatother
\newlength{\thickarrayrulewidth}
\usepackage{tikz}
\usetikzlibrary{positioning}
\usetikzlibrary{shapes.geometric, arrows}
\usetikzlibrary{arrows.meta}
\usepackage{varwidth}
\usepackage{xcolor}
\definecolor{mtplotlib1}{HTML}{1f77b4}
\definecolor{mtplotlib2}{HTML}{ff7f0e}
\definecolor{mtplotlib3}{HTML}{2ca02c}
\definecolor{mtplotlib4}{HTML}{d62728}

\tikzset{%
  >={Latex[width=2mm,length=2mm]},
            base/.style = {rectangle, rounded corners, draw=black,
                           minimum width=4cm, minimum height=1cm,
                           text centered}, 
            mystyle/.style={rectangle, rounded corners, draw=black,
            minimum width=12cm, minimum height=1cm,
            text centered}, 
    col0/.style = {base, fill=white!30},
    col1/.style = {base, fill=mtplotlib1!30},
    col11/.style = {mystyle, fill=mtplotlib1!30},
    col2/.style = {base, fill=mtplotlib2!30},
    col3/.style = {base, fill=mtplotlib3!30},
    col4/.style = {base, minimum width=2.5cm, fill=mtplotlib4!15,}
}

\newcommand{\be}{\begin{equation}}
\newcommand{\ee}{\end{equation}}
\newcommand{\bea}{\begin{eqnarray}}
\newcommand{\eea}{\end{eqnarray}}
\newcommand{\bi}{\begin{itemize}}
\newcommand{\ei}{\end{itemize}}
\newcommand{\ben}{\begin{enumerate}}
\newcommand{\een}{\end{enumerate}}
\newcommand{\la}{\left\langle}
\newcommand{\ra}{\right\rangle}
\newcommand{\lc}{\left[}
\newcommand{\rc}{\right]}
\newcommand{\lp}{\left(}
\newcommand{\rp}{\right)}

\def\frac#1#2{{{#1}\over {#2}}}
\def\gsim{\mathrel{\rlap{\lower4pt\hbox{\hskip1pt$\sim$}}
    \raise1pt\hbox{$>$}}}       
\def\lsim{\mathrel{\rlap{\lower4pt\hbox{\hskip1pt$\sim$}}
    \raise1pt\hbox{$<$}}}

\newcommand{\rep}{\mathrm{rep}}

\newcommand{\draft}[1]{}

\def\beq{\begin{equation}}
\def\eeq{\end{equation}}

\numberwithin{equation}{section}
\numberwithin{figure}{section}
\numberwithin{table}{section}

\usepackage{tabularx}
\newcolumntype{C}[1]{>{\centering\arraybackslash}p{#1}}

\definecolor{darkblue}{rgb}{0.0,0,0.5}
\definecolor{darkgreen}{rgb}{0.0,0.3,0.0}
\definecolor{redish}{rgb}{0.675,0,0.2}
\definecolor{red}{rgb}{0.8,0,0}
\definecolor{green}{rgb}{0,0.6,0}
\definecolor{bluish}{rgb}{0.2,0.2,0.675}
\definecolor{mygrey}{rgb}{0.6,0.6,0.6}

\usepackage{tikz} 
\usetikzlibrary{shapes,arrows,positioning,automata,backgrounds,calc,er,patterns,arrows.meta}
\usepackage{tikz-feynman}
\tikzfeynmanset{compat=1.0.0}
\usepackage{varwidth}
\usepackage{xcolor}
\definecolor{mtplotlib1}{HTML}{1f77b4}
\definecolor{mtplotlib2}{HTML}{ff7f0e}
\definecolor{mtplotlib3}{HTML}{2ca02c}
\definecolor{mtplotlib4}{HTML}{d62728}

\tikzset{%
  >={Latex[width=2mm,length=2mm]},
            base/.style = {rectangle, rounded corners, draw=black,
                           minimum width=4cm, minimum height=1cm,
                           text centered}, 
            mystyle/.style={rectangle, rounded corners, draw=black,
            minimum width=12cm, minimum height=1cm,
            text centered}, 
    col0/.style = {base, fill=white!30},
    col1/.style = {base, fill=mtplotlib1!30},
    col11/.style = {mystyle, fill=mtplotlib1!30},
    col2/.style = {base, fill=mtplotlib2!30},
    col3/.style = {base, fill=mtplotlib3!30},
    col4/.style = {base, minimum width=2.5cm, fill=mtplotlib4!15,}
}

\usetikzlibrary{shapes, arrows}
\usetikzlibrary{decorations.pathreplacing}
\usetikzlibrary{positioning, calc}
\tikzstyle{fitted} = [rectangle, minimum width=5cm, minimum height=1cm, text centered, draw=black, fill=red!30]
\tikzstyle{operations} = [rectangle, rounded corners, minimum width=2cm,text centered, draw=black, fill=red!30]
\tikzstyle{roundtext} = [rectangle, rounded corners, minimum width=2cm, minimum height=0.8cm, text centered, draw=black, fill=red!30]
\tikzstyle{n3py} = [rectangle, rounded corners, minimum width=3cm, minimum height=1cm, text centered, draw=black, fill=green!30]
\tikzstyle{myarrow} = [thick,->,>=stealth]
\tikzstyle{line} =[draw, -latex']
\tikzstyle{decision} = [diamond, draw, fill=red!20, text width=7.5em, text centered,  inner sep=0pt, minimum height=2em, aspect=4]
\tikzstyle{cloud} = [draw, ellipse,fill=green!20, minimum height=2em]
\tikzstyle{inout} = [rectangle, draw, fill=green!20, text width=9.5em, text centered, rounded corners, minimum height=2em, minimum width=10em]
\tikzstyle{block}=[rectangle, draw, fill=blue!20, text width=9.5em, 
                   text centered, rounded corners, minimum height=2em, 
                   minimum width=10em]

\definecolor{darkgreen}{rgb}{0.0, 0.5, 0.13}

\usepackage{tabularx}
\usepackage{subcaption}
\usepackage{pdflscape}
\usepackage{threeparttable}

\newcolumntype{C}[1]{>{\centering\arraybackslash}p{#1}}

\begin{document}
\newgeometry{top=1.5cm,bottom=1.5cm,left=1.5cm,right=1.5cm,bindingoffset=0mm}

\vspace{-2.0cm}
\begin{flushright}
CERN-TH-2024-168\\
\end{flushright}
\vspace{0.6cm}

\begin{center}
  {\Large \bf Hyperparameter Optimisation in Deep Learning from \\[0.2cm] Ensemble Methods: Applications to Proton Structure}\\
  \vspace{1.1cm}
  {\small
Juan Cruz-Martinez$^1$, Aaron Jansen$^2$, Gijs van Oord$^2$,  Tanjona R. Rabemananjara$^{3,4}$, \\[0.1cm] Carlos M. R. Rocha$^2$,  Juan Rojo$^{1,3,4}$, and Roy Stegeman$^{5}$  
  }\\
  
\vspace{0.7cm}

{\it \small
 ~$^1$CERN, Theoretical Physics Department, CH-1211 Geneva 23, Switzerland\\[0.1cm]
 ~$^2$ Netherlands eScience Center, Science Park 140,
1098 XG Amsterdam, The Netherlands \\[0.1cm]
    ~$^3$Department of Physics and Astronomy, Vrije Universiteit, NL-1081 HV Amsterdam\\[0.1cm]
    ~$^4$Nikhef Theory Group, Science Park 105, 1098 XG Amsterdam, The Netherlands\\[0.1cm]
    ~$^5$The Higgs Centre for Theoretical Physics, University of Edinburgh,\\[0.1cm]
   JCMB, KB, Mayfield Rd, Edinburgh EH9 3FD, Scotland\\[0.1cm]
 }

\vspace{1.0cm}

{\bf \large Abstract}

\end{center}

Deep learning models are defined in terms of a large number of hyperparameters, such as network architectures and optimiser settings.
These hyperparameters must be determined separately from the model parameters such as network weights, and are often fixed by ad-hoc methods or by manual inspection of the results.
An algorithmic, objective determination of hyperparameters demands the introduction of dedicated target metrics, different from those adopted for the model training. 
Here we present a new approach to the automated determination of hyperparameters in deep learning models based on statistical estimators constructed from an ensemble of models sampling the underlying  probability distribution in model space.
This strategy requires the simultaneous parallel training of up to several hundreds of models and can be effectively implemented by deploying hardware accelerators such as GPUs. 
As a proof-of-concept, we apply this method to the determination of the partonic substructure of the proton within the NNPDF framework and demonstrate the robustness of the resultant model uncertainty estimates.
The new GPU-optimised NNPDF code results in a speed-up of up to two orders of magnitude, a stabilisation of the memory requirements, and a reduction in energy consumption of up to 90\% as compared to sequential CPU-based model training.
While focusing on proton structure, our method is fully general and is applicable to any deep learning problem relying on hyperparameter optimisation for an ensemble of models. 

\clearpage

\tableofcontents

\section{Introduction}
\label{sec:intro}

Machine Learning (ML) and Artificial Intelligence (AI) techniques are ubiquitous in modern physics~\cite{Carleo:2019ptp,Biamonte:2016ugo}, as acknowledged with the Nobel Prize in Physics 2024.
In the specific case of particle physics at energy frontier facilities such as the Large Hadron Collider (LHC) at CERN in Geneva, ML methods have enabled novel insights and enhanced sensitivities in applications from event classification and anomaly detection in searches for new heavy particles~\cite{Collins:2018epr,CMS:2020poo} to the modelling of jet substructure and heavy-flavour jet tagging~\cite{Larkoski:2017jix,Fraser:2018ieu,Kasieczka:2019dbj}, the design of high-sensitivity observables for Effective Field Theory analyses~\cite{Chen:2020mev,GomezAmbrosio:2022mpm,Brehmer:2018eca,Schofbeck:2024zjo}, and the determination of the substructure of protons~\cite{NNPDF:2014otw,NNPDF:2017mvq} and heavy nuclei~\cite{AbdulKhalek:2020yuc}, among many others.
A more extensive  overview of ML applications to particle physics can be found in the reviews of~\cite{Feickert:2021ajf,Radovic:2018dip,Albertsson:2018maf,annurev:/content/journals/10.1146/annurev-nucl-101917-021019,Bourilkov:2019yoi} and references therein.

Irrespective of their specific implementation, all deep learning models rely on a potentially large number of algorithm-specific hyperparameters, such as the network architecture, the types of layers and connections, the activation functions, the dropout rate, the batch size, the gradient clipping, and the optimiser settings and learning rates.
Furthermore,  different ML architectures come accompanied with their own hyperparamerers, such as the stride, padding, and pooling size present in Convolutional Neural Networks (CNNs) and the number of heads and of encoder/decoder layers defining Transformer models. 
These hyperparameters need to be determined separately from the training of the algorithm-independent parameters, {\it e.g.} the network weights and thresholds, by means of a dedicated, separate procedure.

A wealth of different hyperparameter optimisation methods for applications in machine learning have been put forward, as summarized in the reviews~\cite{DBLP:journals/corr/abs-2003-05689,Feurer2019,YANG2020295,https://doi.org/10.1002/widm.1484} and references therein.
These include techniques such as Bayesian hyperoptimisation~\cite{WU201926}, gradient-based using reversible learning~\cite{pmlr-v37-maclaurin15}, particle swarm optimisation~\cite{Lorenzo2017ParticleSO}, and evolutionary methods~\cite{DBLP:journals/corr/abs-1902-06827}.
Each of these hyperoptimisation algorithms is defined by two main components: first, the metric to be optimised, and second, the strategy to sample the hyperparameter space to find configurations that maximize this sought-for metric. 
Crucially, suitable definitions of the hyperoptimisation metrics depend sensitively on the features that are desirable for a satisfactorily trained ML model, and hence will differ in each application.
This is in contrast with the regular model training, which can in most cases be expressed in terms of a common likelihood maximization problem.

In this work, building upon and extending the results of~\cite{Carrazza:2019mzf,NNPDF:2021njg}, we introduce a novel approach to the automated determination of optimal hyperparameters in deep learning problems.
Our strategy is based on introducing metrics constructed from an ensemble of ML models sampling the underlying probability distribution in model space.
These metrics are hence sensitive to higher moments (e.g. variances and correlations) of the model distribution, and not only to the first one (the mean), as in other available methods including those of~\cite{Carrazza:2019mzf,NNPDF:2021njg}.
Implementing such estimators requires the simultaneous training of ensembles composed of up to several hundreds of models, and also demands scanning efficiently the high-dimensional hyperparameter space to identify the preferred configuration.

We demonstrate the feasibility of the proposed hyperoptimisation approach by means of an application to the determination of proton substructure~\cite{Gao:2017yyd,Ethier:2020way,Kovarik:2019xvh,Amoroso:2022eow}.
As opposed to point-like, structure-less particles such as the electron, the proton is composed by constituents known as partons~\cite{Gao:2017yyd,Kovarik:2019xvh}. These include the valence up and down quarks, giving the proton its quantum numbers, as well as the gluons, the mediators of the strong nuclear force, sea and heavy quarks, photons~\cite{Manohar:2016nzj,Manohar:2017eqh,Bertone:2017bme}, and at high-enough energies even top quarks~\cite{Han:2014nja}, weak gauge bosons, and the Higgs boson itself~\cite{Bauer:2017isx}.
Proton substructure is encoded by objects known as the parton distribution function (PDFs) and cannot be reliably evaluated from first-principle calculations, at least with current technology~\cite{Lin:2017snn,Constantinou:2020hdm}.
For this reason, PDFs need to be extracted from a phenomenological analysis of high-energy data collected in particle colliders, choosing some model for their dependence with the partonic quark flavour and with the partonic momentum fraction $x$.

Within the NNPDF global analysis framework of proton structure~\cite{Ball:2008by, NNPDF:2021uiq}, neural networks are deployed as universal unbiased interpolants to determine the proton PDFs while minimizing theory assumptions.
Complemented with the Monte Carlo ensemble method~\cite{Forte:2002fg,DelDebbio:2004xtd} for uncertainty estimate and propagation, the NNPDF methodology is now well-established and has enabled a number of insights on proton structure, from identifying a gluonic contribution to the proton spin~\cite{Nocera:2014gqa} and the onset of BFKL dynamics in HERA data~\cite{Ball:2017otu} to obtaining  evidence for intrinsic heavy quarks in the proton~\cite{Ball:2022qks} and for gluon shadowing in heavy nuclei~\cite{AbdulKhalek:2022fyi}.
Beyond applications to particle physics, the NNPDF methodology has also been successfully applied to the automated interpretation of Electron Energy-Loss Spectroscopy (EELS) measurements of low-dimensional quantum materials within the {\sc\small EELSfitter} framework~\cite{Roest:2020kqy,Brokkelkamp:2022wjw,ExcitonAnatomy,LA2023113841}.

The current NNPDF methodology is based on hyperparameters obtained from metrics evaluated in terms of the $\chi^2$ to the central dataset, computed on out-of-sample datasets using $K$-folds cross-validation~\cite{Carrazza:2019mzf,NNPDF:2021njg}. 
Here we show how to extend this procedure to more general metrics built upon the full probability distribution in model space as spanned by the Monte Carlo replicas of the trained ensemble, which encode sensitivity to the PDF uncertainties and correlations associated to each model.
Constructing such metrics demands the simultaneous training of hundreds of models simultaneously, which we make possible by extending the NNPDF framework to state-of-the-art hardware accelerators, in particular to Graphical Processing Units (GPUs).
We then apply the developed strategy to carry out a variant of the NNPDF4.0 determination, assess the stability of the results both for central values and for PDF errors, and quantify the resulting improvements in terms of computational efficiency, memory requirements, and energy consumption.

The outline of this paper is as follows.
First, in Sect.~\ref{sec:hyperopt} we recapitulate the hyperparameter selection method used in the NNPDF4.0 analysis and describe the new metrics based on the full replica ensemble.
Sect.~\ref{sec:implementation} highlights the technical implementation of our approach, including the GPU training, as well as the validation checks and performance studies carried out.
The main results of this work are provided in Sect.~\ref{sec:results}, where we present a 
variant of NNPDF4.0 based on the new hyperparameter selection strategy. 
We summarize and discuss possible future developments in Sect.~\ref{sec:summary}.
Additional information on technical improvements is provided in App.~\ref{app:techimpr}.

\section{Hyperparameter optimisation strategies}
\label{sec:hyperopt}

We begin by providing a self-contained introduction to the NNPDF methodology for proton structure analyses.
Then, we review the hyperparameter optimisation strategy that was adopted in the NNPDF4.0 study.
Subsequently, we introduce the new hyperparameter optimisation approach based on metrics constructed in terms of the complete probability density in the space of PDF models. 

\subsection{Proton structure from machine learning}
\label{subsec:nnpdf_summary}

Within the NNPDF fitting methodology, the main outcome~\cite{Forte:2002fg,DelDebbio:2007ee} is a sampling of the probability distribution in the space of PDFs expressed in terms of an ensemble of $N_{\rm rep}$ Monte Carlo replicas,
\be
\label{eq:sampling}
f_\ell^{(k)}(x,Q^2) \,, \quad k=1,\ldots,N_{\rm rep} \,, \quad \ell=1,\ldots,n_{f} \, , 
\ee
with $n_{f} $ indicating the partonic flavour combinations being considered. 
Information on the features of the resultant PDF distribution can be obtained by evaluating suitable estimators over the sampling given by Eq.~(\ref{eq:sampling}), e.g., PDF uncertainties can be computed by taking the standard deviation of the distributions,
\be
\Delta_{\rm pdf}f_\ell(x,Q^2) = \lp \frac{1}{N_{\rm rep}}\sum_{k=1}^{N_{\rm rep}}\lp f_\ell^{(k)}(x,Q^2)  - \la f_\ell(x,Q^2) \ra_{\rm rep}\rp\rp^{1/2} \, .
\ee
Each of the $N_{\rm rep}$ Monte Carlo replicas composing Eq.~(\ref{eq:sampling}) is obtained from a fit to an independent Monte Carlo fluctuation $\{ D^{(k)}\}$ of the original  experimental data, $\{ D\}$, produced accordingly to the associated covariance matrix.
The latter may also account for theoretical uncertainties by means of the theory covariance matrix method~\cite{NNPDF:2019ubu,NNPDF:2019vjt,NNPDF:2024dpb}.
This fit is based on the minimisation of a suitable figure of merit, namely
\begin{equation}
    E^{(k)} = \frac{1}{n_{\rm dat}}\displaystyle\sum_{i,j=1}^{n_{\rm dat}} \lp D^{(k)}_{i}-T^{(k)}_{i} \rp\lp {\rm cov}\rp^{-1}_{ij} \lp D^{(k)}_{j}-T^{(k)}_{j}\rp \,,\label{eq:chi2definition}
\end{equation}
with $\{ D^{(k)}\}$ indicating the $k$-th replica of the experimental data, $\{ T^{(k)}\}$ the associated theoretical predictions, the covariance matrix includes all relevant sources of experimental~\cite{Ball:2009qv} and theory uncertainties and is the same one used to generate the replicas $\{ D^{(k)}\}$, and $n_{\rm dat}$ indicates the total number of datapoints.

The theory predictions entering Eq.~(\ref{eq:chi2definition}) can be expressed in terms of the PDFs $f_i(x,Q^2_0)$ at the input parametrisation scale $Q_0$ ($\sim 1~{\rm GeV}$) in the following manner:
\begin{equation}
\label{eq:predict}
T^{(k)}_{i}=\sum_{m,n=1}^{n_x}\sum_{\ell,\ell'=1}^{n_f}{\rm FK}_{i,mn}^{\ell\ell'}f_{\ell,m}^{{(k)}}(Q_0) 
f_{\ell',n}^{{(k)}}(Q_0)\,,
\end{equation}
for hadron collider observables and
\begin{equation}
\label{eq:predict_dis}
T^{(k)}_{i}=\sum_{m=1}^{n_x}\sum_{\ell=1}^{n_f}{\rm FK}_{i,m}^{\ell}f_{\ell,m}^{{(k)}}(Q_0) \,,
\end{equation}
for deep-inelastic scattering (DIS) observables.
$n_f$ is as above the number of partonic flavours, 
the tensors ${\rm FK}$ are FastKernel (FK) tables~\cite{Ball:2010de},
PDF-independent interpolation grids which encode all relevant information on the partonic hard-scattering cross-section and the DGLAP evolution kernels, such that the observables $\{ T^{(k)}\}$ entering the fit can be evaluated in a computationally efficient manner.
$n_x$ denotes the number of $x$-nodes in the FK Table.
The PDFs are then sampled as
\be
f_{\ell,m}^{{(k)}}(Q_0)\equiv f_{\ell}^{{(k)}}(x_m,Q_0)\,, \quad m=1,\ldots,n_{x} \,, \quad \ell=1,\ldots,n_{f} \,
\ee
where each of the PDF replicas $f_{\ell,m}^{{(k)}}(Q_0)$ define the PDF distribution and are at the same time physically valid PDFs.

Several strategies are possible to generate the FK-tables; in this work we use the methodology adopted in~\cite{NNPDF:2024djq,NNPDF:2024dpb,NNPDF:2024nan} and based on interfacing the grid interpolation library {\sc\small PineAPPL}~\cite{Carrazza:2020gss, christopher_schwan_2024_13951776} with {\sc\small EKO}~\cite{Candido:2022tld, candido_2022_6340153}, see~\cite{Barontini:2023vmr} for more details.

The PDFs  $f_i(x,Q^2_0)$ are parametrised in the NNPDF approach at the input scale $Q_0$ in terms of feed-forward neural networks, which provide a universal interpolation minimising the need for theory assumptions entering the PDF parametrization~\cite{Ball:2016spl}. 
The goal of a NNPDF fit is to determine the weights and biases of the neural network replicas, denoted by $\boldsymbol{\theta}^{(k)}$, such that the resulting PDFs $f_i^{(k)}(x,Q^2_0,\boldsymbol{\theta})$ minimize the loss function Eq.~(\ref{eq:chi2definition}) while preventing overlearning.
In the NNPDF fits the loss function Eq.~(\ref{eq:chi2definition}), containing the contribution from the experimental data, is extended to account also for theoretical constraints, in particular the positivity and integrability of PDFs~\cite{NNPDF:2021njg}.
Therefore, the complete error function to be minimised is
\begin{equation}
    E^{(k)}(\boldsymbol{\theta}) = \frac{1}{n_{\rm dat}}\displaystyle\sum_{i,j=1}^{n_{\rm dat}} \lp D^{(k)}_{i}-T^{(k)}_{i}(\boldsymbol{\theta}) \rp\lp {\rm cov}\rp^{-1}_{ij} \lp D^{(k)}_{j}-T^{(k)}_{j}(\boldsymbol{\theta})\rp 
    +{\rm theory\_constraints(\boldsymbol{\theta})}
    \,,\label{eq:Ekdefinition_2}
\end{equation}
where the second contribution contains Lagrange multipliers that determine their relative strength in the error function as compared to the contribution from experimental data, and whose values are also determined through the hyperoptimisation procedure. 
We note that the error function being minimised, Eq.~(\ref{eq:Ekdefinition_2}) is independent for each replica.
In the following, it is understood that the minimisation of the loss functions Eq.~(\ref{eq:Ekdefinition_2}) includes a regularisation procedure to prevent overlearning such as dynamical stopping. 


\subsection{Hyperparameter optimisation in NNPDF4.0}
\label{sec:hyperopt_nnpdf40}

The hyperoptimisation procedure currently adopted in NNPDF is described in~\cite{Carrazza:2019mzf,NNPDF:2021njg}.
Many of the considered hyperparameters are generic to deep-learning problems, such as the choice of neural network architecture, the method for weight initialisation, the training rate, the clipnorm rate, the stopping criteria, and the optimisation algorithm to be used.
These could be roughly divided into model and optimisation hyperparameters.
Other hyperparameters are instead specific to NNPDF fits and are connected to the underlying nature to the physical problem at hand, such as the strength with which integrability and positivity theory constraints are imposed, or the shape of the PDFs in (and near) small- and large-$x$ extrapolation regions, introducing a third type of ``theory hyperparameters''.
For the remainder of this work and to present the general strategy they will all be considered in the same footing and all these hyperparameters will be denoted collectively 
by $\boldsymbol{\hat{\theta}}$, to be distinguished from the model parameters (e.g. network weights) $\boldsymbol{\theta}$, which are always different for every replica.
Provided a hyperoptimisation loss function $L_{\rm hopt}(\boldsymbol{\hat{\theta}})$ is specified, a hyperparameter scan is carried out with the \texttt{hyperopt} library~\cite{hyperopt,Bergstra_2015} based on a Bayesian optimisation algorithm to identify the best configurations.

The choice of hyperoptimisation loss function $L_{\rm hopt}$ is crucial for the success of the overall procedure, and should ensure that the resulting fits exhibit a set of desirable features which are application-dependent. 
Since the goal of a PDF fit is to find models that minimise the error function Eq.~(\ref{eq:Ekdefinition_2}) when applied to the original central data, one possible choice of hyperoptimisation metric would be setting it equal to the $\chi^2$, that is
\be
\label{eq:hyperopt_1}
L_{\rm hopt}(\boldsymbol{\hat{\theta}}) = \underset{\boldsymbol{\theta} \in {\boldsymbol{\Theta}}}{\text{ min}^*}\lc \chi^2\lp \boldsymbol{\theta},\boldsymbol{\hat{\theta}} \rp \rc \, ,
\ee
where $\boldsymbol{\Theta}$ denotes the space of neural network weights and the $\chi^2$ is defined by comparing the theory prediction to the original experimental data, 
\begin{equation}
    \chi^2(\boldsymbol{\theta}) = \frac{1}{n_{\rm dat}}\displaystyle\sum_{i,j=1}^{n_{\rm dat}} \lp D^{(0)}_{i}-T^{(0)}_{i}(\boldsymbol{\theta}) \rp\lp {\rm cov}\rp^{-1}_{ij} \lp D^{(0)}_{j}-T^{(0)}_{j}(\boldsymbol{\theta})\rp 
    +{\rm theory\_constraints(\boldsymbol{\theta})}
    \,,\label{eq:chi2definition_2}
\end{equation}
where $T^{(0)}_{i}(\boldsymbol{\theta})= \la T^{(k)}_{i}(\boldsymbol{\theta}) \ra_{rep} $, and where the dependence with the hyperparameters $\boldsymbol{\hat{\theta}}$ is left implicit.
In Eq.~(\ref{eq:hyperopt_1}) the $^*$ sign indicates that the minimisation is subject to a regularisation procedure preventing overlearning. 
A hyperoptimisation loss function of the type of Eq.~(\ref{eq:hyperopt_1}) is however not advisable, since it directly interferes with the outcome of the training itself (via the stopping parameters, for example), and it is known to lead to configurations prone to overfitting. 
Most importantly, this type of loss functions consider only best-fit configurations but ignore the spread in model space, which is a key feature of the resulting PDFs (defining their uncertainties) and determines its generalisation power.

Instead of the metric Eq.~(\ref{eq:hyperopt_1}), in NNPDF4.0 and subsequent analyses the hyperoptimisation loss function is evaluated through a $K$-folds cross-validation method, which separates the data being used in the fit from the data being used to select the best hyperparameters.
In this procedure, the \texttt{hyperopt} library generates a large number of hyperparameter configurations $\boldsymbol{\hat{\theta}}$, and each of them is used to produce fits to  subsets of the central experimental data $\{ D^{(0)}\}$.
Specifically, for each point in the hyperparameter space, we run $n_\text{K}$ fits to the central experimental data, where $n_\text{K}$ is the number of subsets (folds) in which the data are being divided. 
In each of these $n_\text{K}$ fits, the $p$-th fold is left out; the remaining  folds are combined in a dataset which is then separated into training and validation in the usual way, such that the patience stopping algorithm can be applied.
The hyperoptimisation loss function is then defined as
\be
\label{eq:hyperopt_2}
L_{\rm hopt} \lp \boldsymbol{\hat{\theta}}\rp =  \frac{1}{n_\text{K}}\sum_{p=1}^{n_{\rm K}}\underset{\boldsymbol{\theta} \in {\boldsymbol{\Theta}}}{\text{ min}^*} \lp \chi^2_p\lp \boldsymbol{\theta},\boldsymbol{\hat{\theta}} \rp  \rp \, ,
\ee
where $\chi^2_p$ indicates the value of the $\chi^2$ error function Eq.~(\ref{eq:chi2definition_2}), with the contribution from the theory constraints excluded, evaluated on the $p$-th fold datapoints $n_p$ which have been left out from the $p$-th fit, 
\begin{equation}
    \chi^2_p(\boldsymbol{\theta}) = \frac{1}{n_p} \displaystyle\sum_{i,j \in p} \lp D^{(0)}_{i}-T^{(0)}_{i}(\boldsymbol{\theta}) \rp\lp {\rm cov}\rp^{-1}_{ij} \lp D^{(0)}_{j}-T^{(0)}_{j}(\boldsymbol{\theta})\rp 
    \,,\label{eq:chi2definition_5}
\end{equation}
Therefore, the hyperoptimisation loss function defined by Eq.~(\ref{eq:hyperopt_2}) corresponds to the average of the $\chi^2$ evaluated for the (non-fitted) $p$-th folds.
To ensure a well-balanced procedure, each fold is representative of the global dataset, both in terms of process type  and of kinematic coverage in the $(x,Q^2)$ plane.
The optimal hyperparameter set ${\boldsymbol{ \hat{\theta}} }^*$ is then selected to be the one that produces the lowest average loss computed using
Eq.~(\ref{eq:hyperopt_2}), that is
\begin{equation}
\boldsymbol{\hat{\theta}}^* = 
\underset{\boldsymbol{\hat{\theta}} \in {\boldsymbol{\hat{\Theta}}}}{\text{arg min}}\lp L_{\rm hopt} \lp \boldsymbol{\hat{\theta}}\rp\rp 
=
\underset{\boldsymbol{\hat{\theta}} \in {\boldsymbol{\hat{\Theta}}}}{\text{arg min}}\left( 
\frac{1}{n_\text{K}}\sum_{p=1}^{n_{\rm K}} \underset{\boldsymbol{\theta} \in {\boldsymbol{\Theta}}}{\text{ min}^*}\lc\chi^2_p\lp \boldsymbol{\theta},\boldsymbol{\hat{\theta}} \rp  \rc  \right) \, .
	\label{eq:hyperoptloss}
\end{equation}

In the NNPDF4.0 study, alternative choices were considered for the hyperoptimisation loss function, such as
\be
\label{eq:Lmax}
L_{\rm hopt} \lp \boldsymbol{\hat{\theta}}\rp = {\rm max}\lp  \underset{\boldsymbol{\theta} \in {\boldsymbol{\Theta}}}{\text{ min}^*}\lp\chi_{1}^2\lp \boldsymbol{\theta},\boldsymbol{\hat{\theta}} \rp \rp ,  \ldots, 
\underset{\boldsymbol{\theta} \in {\boldsymbol{\Theta}}}{\text{ min}^*}\lp\chi_{n_{\rm K}}^2\lp \boldsymbol{\theta},\boldsymbol{\hat{\theta}} \rp \rp
\rp,
\ee
that is, the maximum value of the $n_{\rm K}$ $\chi^2_p$ values evaluated in the $p$-th fold, Eq.~(\ref{eq:chi2definition_5}).
As in the case of Eq.~(\ref{eq:hyperopt_2}), the alternative loss function Eq.~(\ref{eq:Lmax}) favours hyperparameter configurations which are able to extrapolate best (as measured by the fit quality) to non-fitted datasets.
In~\cite{NNPDF:2021njg} it was verified that the two hyperoptimisation loss functions led to equivalent results for the final PDFs.

The main drawback of the two hyperoptimisation loss functions considered in NNPDF4.0, Eqns.~(\ref{eq:hyperopt_2}) and (\ref{eq:Lmax}), is that they only gauge the agreement between the central value of the theory predictions and the experimental data, but neglect the uncertainties in the former, that is, the PDF errors entering those predictions. 
Neglecting PDF uncertainties in the hyperoptimisation loss function, however, was necessary to keep the computational burden required by the  NNPDF4.0 hyperoptimisation procedure to acceptable levels. 
Indeed, extending the loss function to account for PDF uncertainties would have increased the hyperoptimisation computational overhead by a factor $N_{\rm rep}$, since for each hyperoptimisation trial one fits to the full $N_{\rm rep}$ set of data replicas and not only to the experimental data.

Nevertheless, the resulting PDF uncertainties in  NNPDF4.0 obtained using the hyperoptimisation methodology described here were thoroughly validated {\it a posteriori} by verifying~\cite{NNPDF:2021njg} that {\it i)} PDF uncertainties are correctly estimated in the case of a known underlying theory, the so-called closure tests~\cite{NNPDF:2014otw,DelDebbio:2021whr}, and that {\it ii)} these PDFs can be extrapolated to predict, within uncertainties, experiments not included in the fit and whose kinematics belong to the (deep) extrapolation regions, the so-called future tests~\cite{Cruz-Martinez:2021rgy}. 

\subsection{Ensemble-based hyperoptimisation}
\label{sec:hyperopt_new}

The hyperoptimisation metrics introduced in Sect.~\ref{sec:hyperopt_nnpdf40} are only sensitive to the central value of the theory predictions, but not to higher moments of the distribution such as the PDF uncertainties (variances). 
To bypass this limitation, here we extend the NNPDF framework to carry out multi-replica parallel fits, where the $N_{\rm rep}$ replicas defining a PDF set are trained simultaneously while decreasing the overall computational overhead.
This powerful feature enables us to define novel hyperoptimisation metrics that are sensitive to the PDF uncertainties, and eventually to higher moments of the probability distribution, leading to a more robust determination of the model hyperparameters $\boldsymbol{\hat{\theta}}$. 

\paragraph{First-moment metrics.}
When developing hyperoptimisation metrics evaluated over the model probability distribution, the simplest option would be to use the same metric as in NNPDF4.0, Eq.~(\ref{eq:hyperopt_2}), but now extended to the $\chi^2$ computed from the average over replicas rather than limited to the central value.
In this scenario, one may define a metric such as
\be
\label{eq:hyperopt_2_replicas}
L_{\rm hopt}^{(\chi^2)} \lp \boldsymbol{\hat{\theta}}\rp =  \frac{1}{n_\text{fold}}\sum_{p=1}^{n_{\rm fold}}\underset{\boldsymbol{\theta} \in {\boldsymbol{\Theta}}}{\text{ min}^*} \lp \ \la \chi^2_p\lp \boldsymbol{\theta},\boldsymbol{\hat{\theta}} \rp \ra_{\rm rep} \rp \, ,
\ee
where instead of Eq.~(\ref{eq:chi2definition_5}), the $\chi^2_p$ for the $p$-th fold is evaluated over the replicas and then averaged out,
\begin{equation}
  \la \chi^2_p\lp \boldsymbol{\theta},\boldsymbol{\hat{\theta}} \rp \ra_{\rm rep} =  \frac{1}{N_{\rm rep}} \sum_{k=1}^{N_{\rm rep}}
  \chi^{2(k)}_p\lp \boldsymbol{\theta},\boldsymbol{\hat{\theta}} \rp \, ,
  \label{eq:chi2p_average_over_replicas}
\ee
where
\begin{equation}
    \chi^{2(k)}_p(\boldsymbol{\theta}) = \frac{1}{n_p} \displaystyle\sum_{i,j \in p} \lp D^{(0)}_{i}-T^{(k)}_{i}(\boldsymbol{\theta}) \rp\lp {\rm cov}\rp^{-1}_{ij} \lp D^{(0)}_{j}-T^{(k)}_{j}(\boldsymbol{\theta})\rp 
    \,,\label{eq:chi2definition_6}
\end{equation}
is restricted to the datapoints which exclude the $p$-th data fold, as discussed in Sect.~\ref{sec:hyperopt_nnpdf40}.
Results of hyperparameter optimisation based on the metric Eq.~(\ref{eq:hyperopt_2_replicas}) would be expected to be similar to those based on Eq.~(\ref{eq:hyperopt_2}), since in the former the information on higher moments of the distribution is partially washed out when averaging of the replicas.
For this reason, Eq.~(\ref{eq:hyperopt_2_replicas}) can provide a baseline to compare the performance of other hyperoptimisation metrics based on higher moments of the full probability distribution.

Before evaluating Eq.~(\ref{eq:hyperopt_2_replicas}), one should ensure that the replicas entering the average in Eq.~(\ref{eq:chi2p_average_over_replicas}) satisfy the usual set of post-fit quality criteria imposed in the NNPDF fits~\cite{NNPDF:2014otw,NNPDF:2021njg}.
There one trains a larger number of replicas than the strict minimum required, and then rejects outliers that contaminate the PDF distribution.
These outliers can arise, for instance, in the case of replicas with features deemed to be unphysical like rapid oscillations.
This outlier filtering procedure, denoted as {\tt postfit} in the NNPDF framework, is based on evaluating variances over the relevant distributions, e.g. $\chi^{2(k)}$, and then discarding those replicas that are several sigma away from the mean value of the distribution and which are hence deemed to be statistically unlikely given the sample size.

In this work, when evaluating the averages in Eq.~(\ref{eq:chi2p_average_over_replicas}), we adopt a computationally cheaper quality selection criterion based on removal of a fraction $(1-f_{\rm filter})$ of the replicas with the largest $\chi^{2(k)}_p$ values, such that
\be
\label{eq:hyperopt_2_replicas_postfit}
  \la \chi^2_p\lp \boldsymbol{\theta},\boldsymbol{\hat{\theta}} \rp \ra_{\rm rep,postfit} =  \frac{1}{f_{\rm filter}N_{\rm rep}} \sum_{k=1}^{f_{\rm filter}N_{\rm rep}}
  \chi^{2(k)}_{p,{\rm sorted}}\lp \boldsymbol{\theta},\boldsymbol{\hat{\theta}} \rp \, ,
\ee
with $\chi^{2(k)}_{p,{\rm sorted}}$ indicating that the per-replica $\chi^{2(k)}_p$ values have been sorted in order of increasing value.
It is found that a choice of $f_{\rm filter}=0.9$ is appropriate to achieve stability with respect to outliers and to remove poorly-convergent replicas.

\paragraph{Improved first-moment metrics.}
One drawback of Eq.~(\ref{eq:hyperopt_2_replicas}) is that hyperparameter selection may be driven by a relatively small subset of data points which are far from the theory prediction in units of the data uncertainty.
The relative data vs theory weighting would be different if one were to also account for PDF uncertainties.
In other words,  Eq.~(\ref{eq:hyperopt_2_replicas}) is only a suitable figure of merit in cases where experimental errors are much larger than PDF errors.
A better definition would be one which measures the deviation of the central value of the theory prediction with respect to the data in units of the combined experimental and PDF uncertainties.

With this motivation, an improved version of Eq.~(\ref{eq:hyperopt_2_replicas}) is defined as 
\be
\label{eq:hyperopt_2_replicas_PDF}
L_{\rm hopt}^{(\chi^2_{\rm pdf})} \lp \boldsymbol{\hat{\theta}}\rp =  \frac{1}{n_\text{fold}}\sum_{p=1}^{n_{\rm fold}}\underset{\boldsymbol{\theta} \in {\boldsymbol{\Theta}}}{\text{ min}^*} \lp \ \la \chi^2_{{\rm PDF},p}\lp \boldsymbol{\theta},\boldsymbol{\hat{\theta}} \rp \ra_{\rm rep} \rp \, ,
\ee
where now the $\chi^{2(k)}_{p}$ in the $p$-th fold includes the contribution from the PDF uncertainties, added in quadrature to the experimental covariance matrix
\begin{equation}
    \chi^{2(k)}_{{\rm PDF},p}(\boldsymbol{\theta}) = \frac{1}{n_p} \displaystyle\sum_{i,j \in p} \lp D^{(0)}_{i}-T^{(k)}_{i}(\boldsymbol{\theta}) \rp\lp {\rm cov}^{\rm (exp)} + {\rm cov}^{\rm (PDF)}  \rp^{-1}_{ij} \lp D^{(0)}_{j}-T^{(k)}_{j}(\boldsymbol{\theta})\rp 
    \,,\label{eq:chi2definition_7}
\end{equation}
following the theory covariance matrix formalism~\cite{NNPDF:2024dpb}.
While Eq.~(\ref{eq:hyperopt_2_replicas_PDF}) can only be computed a posteriori, once the PDF ensemble is available, it is more stable with respect to data points with small uncertainties, since now the distance between theory and data also accounts for the PDF errors to determine the goodness-of-fit,
and is adopted in the following analysis.

\paragraph{Second-moment metrics.}
The main limitation of the hyperoptimisation metric Eq.~({\ref{eq:hyperopt_2_replicas_PDF}) is that it is only sensitive to the first moment of the distribution and thus neglects information contained in higher moments, in particular concerning the PDF uncertainties.
This restriction can be bypassed by considering the so-called $\varphi_{\chi^2}$ estimator introduced in~\cite{NNPDF:2014otw} and defined as
\be
\label{eq:chi_estimator}
\varphi^2_{\chi^2} \equiv \la \chi^2\lc T,D^{(0)}\rc\ra_{\rm rep} - \chi^2 \lc \la T\ra_{\rm rep},D^{(0)}\rc  \, ,
\ee
where, as in Eq.~(\ref{eq:chi2definition_2}), $\{ D^{(0)}\}$ indicates the original experimental data and $\{ T^{(k)}\}$ is the best-fit theory prediction obtained for the $k$-th replica.
Only the contribution from experimental data (and not those associated to the theory constraints) is considered in the definition of $\varphi^2_{\chi^2}$.
The first term in the RHS of Eq.~(\ref{eq:chi_estimator}) coincides with the average over Eq.~(\ref{eq:chi2definition_6}) when evaluated on the full dataset.

The motivation to adopt Eq.~(\ref{eq:chi_estimator}) is that this estimator provides a measure of the standard deviation over the replica sample in data space, that is, the PDF uncertainty associated to the theory predictions for the fitted cross-sections $\{T\}$ in units of the data uncertainty.
Indeed, it can be shown~\cite{NNPDF:2014otw} that one can express this estimator in terms of the covariance matrix and the ensemble of theory predictions:
\be
\label{eq:varphi_estimator_2}
\varphi^2_{\chi^2} = \frac{1}{n_{\rm dat}}\sum_{i,j=1}^{n_{\rm dat}} \lp \rm cov\rp^{-1}_{ij}T_{ji} \, ,
\ee
with the  covariance matrix of the theoretical predictions given by
\be
T_{ij} \equiv \lp \la T_i T_j\ra_{\rm rep} -\la T_i \ra_{\rm rep}
\la T_j \ra_{\rm rep}
\rp \, .
\ee
Eq.~(\ref{eq:varphi_estimator_2}) highlights how the $\varphi^2_{\chi^2}$ estimator corresponds to the average, over the fitted data points, of the uncertainties and correlations of the theoretical predictions normalized according to their counterparts at the level of the experimental measurements.
In other words, Eq.~(\ref{eq:varphi_estimator_2}) measures the PDF uncertainties in units of the data uncertainties for the experimental observables entering the fit.

Following the reasoning of Sect.~\ref{sec:hyperopt_nnpdf40}, it is important to evaluate the hyperoptomization metrics in terms of data not used in the model training. 
With this motivation, for the $p$-th data fold (whose data points have been excluded from the fit), we define
\be
\label{eq:chi_estimator_pfold}
\varphi^2_{\chi^2_p} \equiv \la \chi^2_p\lc T,D^{(0)}\rc\ra_{\rm rep} - \chi^2_p \lc \la T\ra_{\rm rep},D^{(0)}\rc  \, ,
\ee
with $\chi^{2(k)}_p$ defined in Eq.~(\ref{eq:chi2definition_6}) and evaluated over the data points in the $p$-th fold.
In the context of hyperparameter optimisation, the choice of Eq.~(\ref{eq:chi_estimator_pfold}) enables to select models based on their generalisation properties to the non-fitted $p$-th folds in terms of the PDF uncertainties, and not only in terms of central values (best fit) as previously done.

Indeed, given an ensemble of models describing  satisfactorily the data of the non-fitted folds, one would like to select those which maximise PDF uncertainties in the same non-fitted cross-sections.
A suitable hyperoptimisation loss function displaying this property can be defined as
\be
L_{\rm hopt}^{(\varphi^2)} \lp \boldsymbol{\hat{\theta}}\rp  \equiv \lp  \frac{1}{n_\text{K}}
  \displaystyle\sum^{n_\text{K}}_{p=1} \varphi_{\chi^2_p}^2 \lp \boldsymbol{\hat{\theta}}\rp\rp^{-1} \, ,
\label{eq:hyperoptloss_l3}
\ee
which then selects the hyperparameters leading to the largest PDF errors (in units of the experimental data errors) in the data points excluded from the fit within the $K$-fold procedure.

It should be clear however that the stand-alone application of Eq.~(\ref{eq:hyperoptloss_l3}) as hyperoptimisation loss function  will in general not produce a satisfactory PDF model selection.
The reason is that we don't want to only select models based on the largest PDF uncertainties in the non-fitted folds: we also need to ensure good agreement with the central predictions, as achieved via the metric Eq.~(\ref{eq:hyperopt_2_replicas_PDF}).
Therefore, a successful hyperparameter optimisation procedure in  context of a PDF determination should combine Eq.~(\ref{eq:hyperoptloss_l3}) with Eq.~(\ref{eq:hyperopt_2_replicas_PDF}): the goal is to achieve a satisfactory description of non-fitted data points while favouring those models with the largest PDF uncertainty.

\paragraph{Combined hyperoptimisation algorithm}
To achieve the sought-for features of the new hyperparameter optimisation algorithm, we combine Eq.~(\ref{eq:hyperoptloss_l3}) with Eq.~(\ref{eq:hyperopt_2_replicas_PDF}) into a so-called ``minimum $\chi^2$ maximum $\varphi^2_{\chi^2}$'' algorithm.
This strategy is composed of the following steps:

\begin{enumerate}[label={\#\arabic*.}]
    
\item Determine the point $\boldsymbol{\hat{\theta}}^*_1$ in the hyperparameter space ${\boldsymbol{\hat{\Theta}}}$ which minimises the first-moment hyperoptimisation metric defined in Eq.~(\ref{eq:hyperopt_2_replicas_PDF}):
    \begin{equation}
    \boldsymbol{\hat{\theta}}_1^* = \underset{\boldsymbol{\hat{\theta}} \in {\boldsymbol{\hat{\Theta}}}}{\text{arg  min}} \left[ L_{\rm hopt}^{(\chi^2_{\rm pdf})} \lp \boldsymbol{\hat{\theta}}\rp \right] \, ,
    	\label{eq:hyperoptloss_2}
    \end{equation}
    and hence we define the reference value of this metric to be
    \begin{equation}
    L_{\rm hopt,min}^{(\chi^2_{\rm pdf})} \equiv 
   L_{\rm hopt}^{(\chi^2_{\rm pdf})} \lp \boldsymbol{\hat{\theta}^*_1}\rp \, .
    \end{equation}

\item{Evaluate the standard deviation $\sigma_{\chi^2}$ from the spread of $\chi_{{\rm PDF},p}^{2(k)}$, Eq.~(\ref{eq:chi2definition_7}), among the $N_{\rm rep}$ replicas of the selected fit with hyperparameters $\boldsymbol{\hat{\theta}}^*_1$,} 

\begin{equation}
    \sigma^2_{\chi^2} = \frac{1}{n_{\rm K}N_{\rm rep}} \sum^{n_\text{fold}}_{p=1} \sum_{k=1}^{N_{\rm rep}}
    \left( \chi^{2(k)}_{{\rm PDF},p} -  L_{\rm hopt,min}^{(\chi^2_{\rm pdf})} \right)^{2} ,
    \label{eq:sigma}
    \end{equation}

The variance evaluated according to Eq.~(\ref{eq:sigma}) provides a measure of the fluctuations of the replicas composing the fit with hyperparameters  $\boldsymbol{\hat{\theta}}^*_1$ around the minimum value of Eq.~(\ref{eq:hyperopt_2_replicas_PDF}).

\item{Select the hyperparameter sets $\boldsymbol{\hat{\theta}}$ that will be used to determine the final ensemble of replicas. 
This selection takes into account both the value of the metrics Eq.~(\ref{eq:hyperopt_2_replicas_PDF}), sensitive to central values, and of Eq.~(\ref{eq:hyperoptloss_l3}), sensitive to PDF uncertainties.}

 In this step, we consider that points $\boldsymbol{\hat{\theta}}$ in the parameter space which satisfy
 \be
L_{\rm hopt}^{(\chi^2_{\rm pdf})} \lp \boldsymbol{\hat{\theta}}\rp \in \lc L_{\rm hopt,min}^{(\chi^2_{\rm pdf})},  L_{\rm hopt,min}^{(\chi^2_{\rm pdf})}+\sigma_{\chi^2} \rc \, ,
\label{eq:step3_combined_algorithm}
 \ee
 provide an equally satisfactory description of the central value of the data in the non-fitted folds. 

In order to select among the hyperparameters $\boldsymbol{\hat{\theta}}$ that comply with Eq.~(\ref{eq:step3_combined_algorithm}), 
we evaluate Eq.~(\ref{eq:hyperoptloss_l3}) on these same $\boldsymbol{\hat{\theta}}$ and then  choose among one of different options:

\begin{itemize}

\item Select a single set of hyperparameters for all replicas from on the minimum value of the metric:

\be
 \boldsymbol{\hat{\theta}}^*_2 \equiv \underset{\boldsymbol{\hat{\theta}} \in {\boldsymbol{\hat{\Theta}}}}{\text{arg  min}} \left[ L_{\rm hopt}^{(\varphi^2)} \lp \boldsymbol{\hat{\theta}}\rp \rc \, .
\ee

\item Identify the $n_{\rm hopt}$ sets of $\boldsymbol{\hat{\theta}}$, satisfying Eq.~(\ref{eq:step3_combined_algorithm}), with the lowest value of  $L_{\rm hopt}^{(\varphi^2)} ( \boldsymbol{\hat{\theta}})$. Each replica will be trained with a set $\boldsymbol{\hat{\theta}}$ uniformly sampled from these $n_{\rm hopt}$ sets.

\item Assign a weight for all the models satisfying Eq.~(\ref{eq:step3_combined_algorithm}) based on their value of the hyperoptimisation metric Eq.~(\ref{eq:hyperoptloss_l3}) with a suitable normalisation e.g.
\be
\label{eq:exponential_weighting}
\omega_\ell \equiv \exp\lp -L_{\rm hopt}^{(\varphi^2)} \lp \boldsymbol{\hat{\theta}_\ell}\rp \Big/ 
 L_{\rm hopt}^{(\varphi^2)} \lp \boldsymbol{\hat{\theta}^*}\rp
\rp \, ,
\ee
and then adopt an unweighting procedure~\cite{Ball:2011gg} such that each replica is trained with a newly sampled set of hyperparameters, possibly with repetition, with a frequency determined by the weights in Eq.~(\ref{eq:exponential_weighting}).

\end{itemize}

\end{enumerate}

For the results presented in Sect.~\ref{sec:results} we will adopt the second option with $n_{\rm hopt}=10$, hence selecting the 10 sets of $\boldsymbol{\hat{\theta}}$ satisfying Eq.~(\ref{eq:step3_combined_algorithm}) with the lowest value of  $L_{\rm hopt}^{(\varphi^2)}$ and then training each Monte Carlo replica with a different set of hyperparameters, sampled uniformly among the selected models.

It should be pointed out that in addition to the use of a different metric for hyperparameter optimisation, a very important difference between the new algorithm introduced here and the one used in NNPDF4.0 is that in the latter all replicas are fitted using exactly the same set of hyperparameters, while now each replica in the fit will be defined by a different set of hyperparameters.
Given the large dimensionality of the hyperparameter space in most ML applications, this procedural modification should result into a more robust estimate of the final PDF uncertainties.

\section{Implementation}
\label{sec:implementation}

Here we describe the implementation of the hyperparameter optimisation strategy described in Sect.~\ref{sec:hyperopt_new}, which requires being able to train PDF replicas in parallel, and quantify the resulting  performance improvements with respect to the NNPDF4.0 baseline.
Additional technical details are provided in App.~\ref{app:techimpr}.

\paragraph{Sequential NNPDF fits.}
The current NNPDF framework is based on training Monte Carlo neural network replicas in a fully independent manner.
The outcome of the training of an individual PDF replica will depend on the choice of random seeds that determine the initialization of the neural network weights and preprocessing exponents, the training/validation mask, and  the data replica generation starting from the underlying experimental measurements.

Being based on training independent replicas, the NNPDF strategy is particularly amenable to leveraging parallel execution capabilities in large computing clusters, provided the number of computing nodes available is greater than the number of replicas requested, the total production time of a PDF fit is determined by the execution time of a single replica, with  different replicas being run as separate concurrent jobs on the common compute infrastructure.
However, it should be emphasized that, in terms of execution and performance, running these multi-replica fits in parallel is formally equivalent to training replicas sequentially one after the other: at no point the workflow is optimised to exploit common aspects of genuine parallel multi-replica training.

For this reason, the current NNPDF workflow exhibits a number of disadvantages, some of which also represent a bottleneck for hyperoptimisation.
The main limitations are:
\begin{itemize}
    \item On busy computing clusters, studies that involve fits with a very large number of replicas, such as those relevant for theory parameter determinations using the correlated replica method of~\cite{Ball:2018iqk}, may lead to long overall delays in the job queue associated to specific single replicas.
    
    \item Within a PDF fit, the most computationally intensive part of the calculation is the convolution (matrix multiplication) of the PDFs with the partonic cross section, Eqns.~(\ref{eq:predict}) and~(\ref{eq:predict_dis}) for hadronic and DIS observables respectively.
    Given that this operation applies to all replicas and does not depend on the neural network's parameters the results for all replicas can be obtained through a single execution. In the current workflow the operation is executed independently for each replica, leading to an inefficiency.

    \item Data loading and other preparation tasks are repeated for each replica, which makes the current NNPDF workflow method rather (energy-) expensive and very intensive on the storage systems (FK Tables can go up to 1 GB).
    
    \item Within the current workflow, the eventual integration of a multi-replica NNPDF fit into a more complex workflow, such as that involved in hyper-optimisation, is tedious and would require non-trivial interactions with the job dispatch and file systems.
    In addition, while individual replicas are trained separately, the total waiting time per hyperparameter configuration is, again, that of the slowest replica.
    This makes adopting an hyperoptimisation strategy based on the ensemble of results, such that the approach of Sect.~\ref{sec:hyperopt_new}, impracticable from a computational point of view unless the workflow is significantly improved.
\end{itemize}

As a prerequisite to realise our new approach to hyperoptimisation, we have therefore significantly improved the NNPDF workflow in order to be able to carry out genuinely parallel multi-replica fits.
The resulting performance advantages also hold for regular NNPDF fits, enabling their deployment in modern hardware accelerators such as GPUs.

\paragraph{Parallel  multi-replica NNPDF fits.}
Whereas some of the limitations of the current NNPDF workflow listed above can be bypassed by brute force, provided the availability of a sufficient amount of computing resources, the ensemble-based hyperoptimisation of Sect.~\ref{sec:hyperopt_new} is unfeasible without a multi-replica version of the PDF fit realising genuine parallel training.
We have thus implemented and optimised a multi-replica parallel training procedure in NNPDF, which allows to seamlessly produce multiple Monte Carlo replicas which are exactly equivalent to the result of the ``sequential'' method used until now in NNPDF.
This result brings in a significant reduction of both of the overall execution time and of the memory footprint, both key requirements to realise ensemble-based hyperoptimisation, with a marked reduction of the energy consumption as by-product. 

The key development is the extension of the neural network architecture used in NNPDF to accommodate the simultaneous training of a large number of parallel replicas.
This way, we effectively increase the problem size to the point for which there is  a clear payoff in utilising the much superior computing power and efficiency of GPUs.
This extension has been achieved by stacking the neural networks that represent each of the Monte Carlo replicas into a single larger model before the convolution with the 
FK tables, Eqns.~(\ref{eq:predict})--(\ref{eq:predict_dis}), and then separating them again immediately afterwards for the training/validation splitting, as indicated by the schematic in Fig.~\ref{fig:stacked_replicas}. 
With this strategy, both the forward and backward passes for the entire ensemble of neural networks required by the back-propagation algorithm are carried out just once per training epoch. 

In order to unify the training procedure of all replicas that are being run in parallel (see Fig.~\ref{fig:stacked_replicas}), it is necessary to adapt the NNPDF methodology as follows.
We define an alternative error function for the simultaneous fit of all replicas as:
\begin{equation}
    \widetilde{E}^{(k)} = \frac{1}{n_{\rm dat}}\displaystyle\sum_{i,j=1}^{n_{\rm dat}} \lp M_{i}^{(k)}\lp  D^{(k)}_{i}-T^{(k)}_{i}\rp \rp\lp {\rm cov}\rp^{-1}_{ij} \lp M_{j}^{(k)}\lp  D^{(k)}_{j}-T^{(k)}_{j} \rp\rp +{\rm theory\_constraints}(\boldsymbol{\theta}^{(k)})\,,\label{eq:chi2definition_masking}
\end{equation}
where both the experimental data and theory predictions are corrected by a replica-specific mask $M_{i}^{(k)}$.
This mask determines whether each data point should be considered as part of  the training or instead of the validation dataset, such that its application effectively recovers the original per-replica error function of Eq.~\eqref{eq:chi2definition} and the stopping algorithm can freeze each independent model separately, making each replica independent of all others such that the result of the fit remains unchanged.
%

In NNPDF4.0, the training/validation masking of the theory predictions is established at initialization by means of a masked version of the FK tables  entering Eqns.~(\ref{eq:predict})--(\ref{eq:predict_dis}). 
This approach has the advantage that memory consumption is reduced and that tensor contractions are evaluated over smaller arrays.
For the multi-replica neural network architecture introduced in this work, we have chosen instead to introduce a separate masking layer built into the network, such that the convolutions with the FK Tables are evaluated jointly for all replicas, with results $T_{i}^{(k)}$ being masked with the tensor $M_{i}^{(k)}$ entering Eq.~(\ref{eq:chi2definition_masking}), see also Fig.~\ref{fig:stacked_replicas}. 
This choice has the key benefit that the operations to be run, in particular the convolutions with the FK Tables, are common across all replicas and can therefore be efficiently run in GPUs at the cost of a somewhat larger greater memory consumption (around 20\%) as compared to single-replica fits.

\begin{figure}[t]
    \centering
    \includegraphics[width=\linewidth]{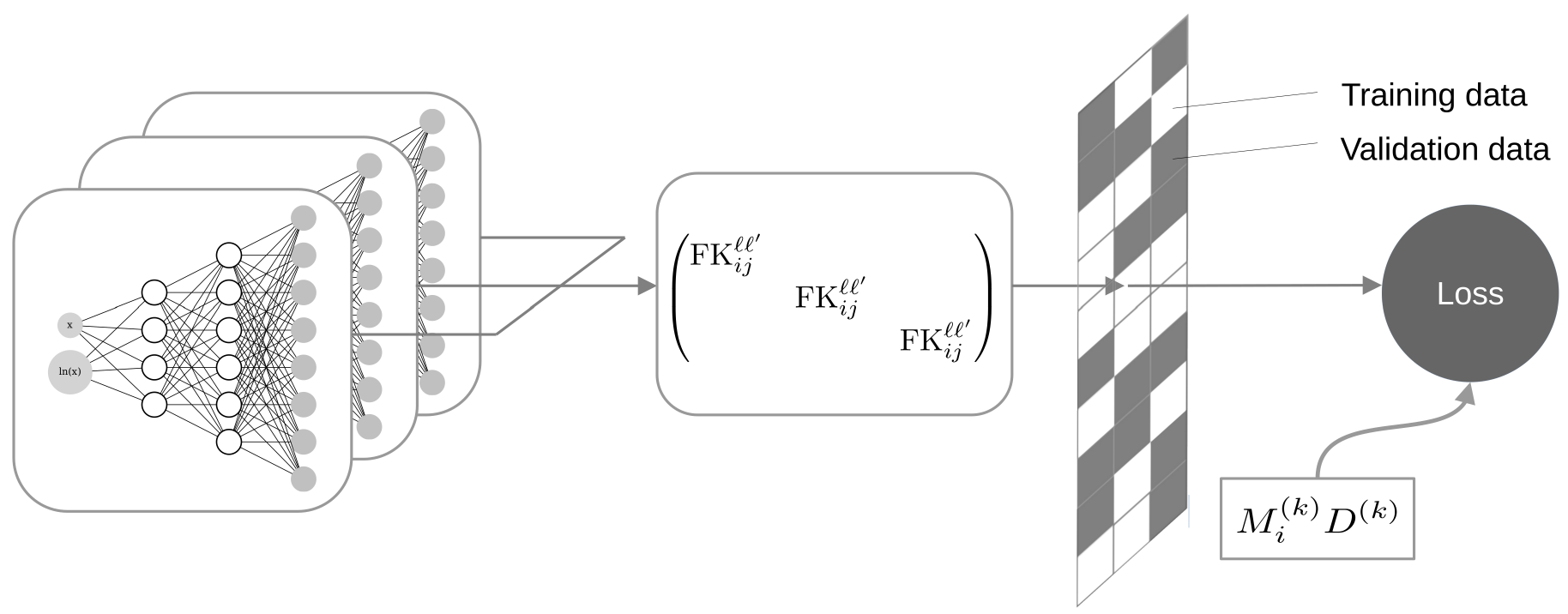}
    \caption{Schematic representation of the model architecture developed here and accommodating multiple stacked replicas.
    The computationally expensive convolution with the FK tables is now shared among all replicas, with the per-replica separation between training and validation data applied immediately afterwards
    through the mask $M_{i}^{(k)}$.
    }
    \label{fig:stacked_replicas}
\end{figure}

The unified calculation of the losses for multiple replicas introduces a new dimension into the problem, and correspondingly exposes new performance bottlenecks in the fitting code. With a few adaptions of the data layout and tensor contraction ordering (see App.~\ref{app:techimpr} for more details), we were able to substantially increase the speed of the multi-replica training workflow.

\paragraph{Performance and benchmarking.}
These technical improvements aimed at optimising the code to benefit from the parallel (Monte Carlo) nature of the problem allow us to take advantage of the performance leaps offered by modern GPUs for AI model training and inference.
For instance, training $\sim 60$ NNPDF replicas simultaneously with this new implementation on a Nvidia H100 GPU is now as fast as the fit of one single replica running on a AMD EPYC Genoa CPU using the original NNPDF4.0 code. 
This significant performance boost provides a feasible pathway towards multi-replica hyperoptimisation with a substantial replica multiplicity, and is the basis of the results presented in Sect.~\ref{sec:results}.

To quantify systematically these performance improvements, Fig.~\ref{fig:multirepbenchmark} (left) displays the scaling of the training speed in the GPU-based multi-replica fits, measured in units of total trained replica models per hour, as a function of the number of parallel replicas $N_{\rep}$, before and after GPU optimisations utilising the settings of NNPDF4.0.
For reference, a dotted line with the sequential training rate on the CPU-based fits is included.
We observe that the GPU optimisations of the FK-table contraction (among other improvements) lead to a factor 2 speed-up as compared to the original NNPDF4.0 code running on GPUs. 
Overall, for a typical ensemble size of $N_{\rm rep}=100$, the achieved speed-up with the new GPU-optimised code is two orders of magnitude as compared to running the NNPDF4.0 baseline code on CPUs.

Then Fig.~\ref{fig:multirepbenchmark} (right) displays the peak memory usage in the same GPU-based fits shown in the left panel as a function of the number of trained replicas $N_{\rep}$ before and after the code optimisations. 
While the increasing chunk size of the TensorFlow memory pool is clearly visible in the original NNPDF4.0 implementation, the FK-table optimisation realised in this work stabilises peak memory usage as $N_{\rm rep}$ increases.
To illustrate this, we checked that for $N_{\rm rep}=100$ replicas the new GPU-optimised NNPDF code requires only half of the 18GB of GPU memory needed by the original NNPDF4.0 code.
From the point of view of multireplica fits, this stabilization of the RAM memory consumption is extremely important, as it allows to run for the first time complete fits in consumer-grade GPUs.
For these two benchmarks, CPU-based fits were executed on a 96-core AMD Genoa 9654 whereas the GPU-based training was done on Nvidia Hopper H100 cards, in both cases running at the national supercomputer {\sc\small Snellius}.
%

Complementing the training speed-up and the decrease in memory requirements,  the parallel GPU-resident multi-replica NNPDF fits developed in this work also lead to a significant reduction in both energy consumption and overall costs in comparison with CPU-based NNPDF4.0 fits, quantified in Table~\ref{tab:energycostcompare}. 
Using the energy-aware runtime technology available on {\sc\small Snellius}, we can monitor power consumption of compute nodes during the model training. 
As expected, the Hopper GPU is more energy-efficient than regular CPU cores, especially
towards larger $N_{\rm rep}$, reaching an energy consumption reduction of 91\% for $N_{\rm rep}=100$.
Similar considerations hold for the cost, where a reduction of 55\% in the GPU-optimised fits as compared to the CPU baseline is found, even taking into account the higher cost of GPU nodes.

\begin{table}[h]
\centering
\small
\renewcommand{\arraystretch}{1.60}
\begin{tabularx}{0.6\textwidth}{lXXXX} 
\toprule
\# Replicas && $\quad$10 & 50 & 100\\ 
\midrule
Energy reduction && $\quad78$\% & 87\% & 91\% \\
Cost reduction && $\,-45$\% & 47\% & 55\% \\
\bottomrule
\end{tabularx}
\vspace{0.3cm}
\caption{Reduction in the energy consumption and on  monetary costs achieved with the parallel GPU-resident multi-replica NNPDF fits developed in this work, when running on {\sc\small Snellius} on a single NVIDIA H100 GPU.
The reference is a pair of concurrent sequential NNPDF fits running on 16 AMD EPYC Genoa CPU cores on {\sc\small Snellius}. 
For the cost calculation we adopt a conversion factor of 196 core-hours per hour per H100 GPU.}
\label{tab:energycostcompare}
\end{table}

\begin{figure}[t]
    \centering
\includegraphics[width=0.49\linewidth]{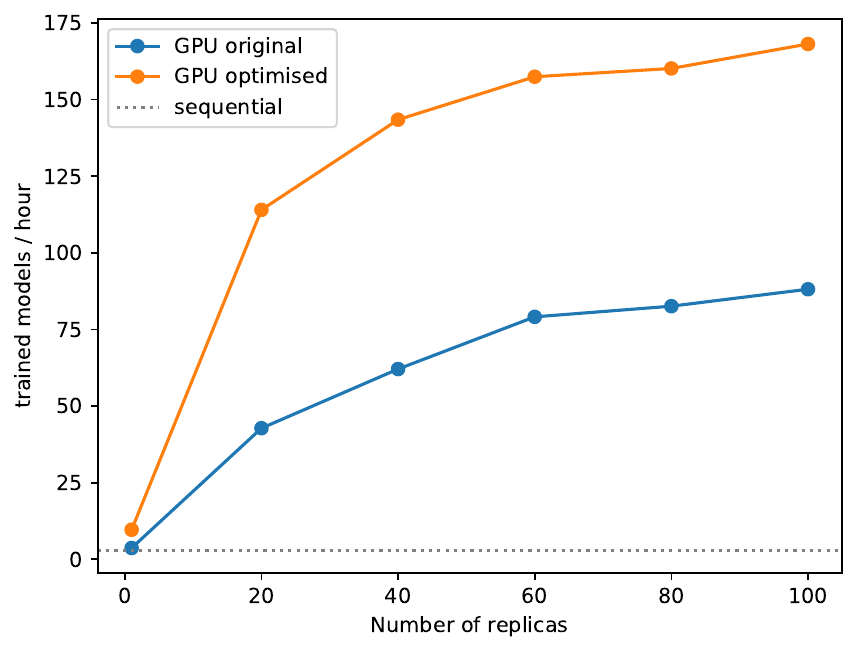}    
\includegraphics[width=0.49\linewidth]{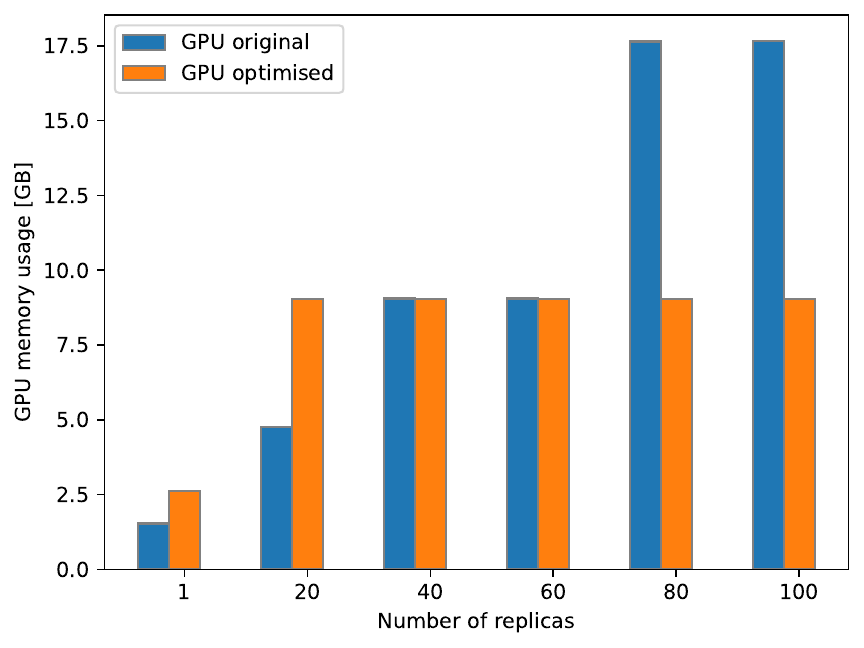}
    \caption{Left: The scaling of the overall training speed in the multi-replica fits developed for this work, measured in units of the number of replica models trained per hour.
    The blue points denote the performance of multi-replica fits on the GPU without optimisations, the orange points are the result of various optimisations described in Appendix \ref{app:techimpr}.
    Right: The peak memory usage (in GB) 
    associated to the fits displayed in the left panel.
    \label{fig:multirepbenchmark}
}
\end{figure}

\paragraph{Parallel GPU-based hyperoptimisation.}

Within the NNPDF framework, the hyperoptimisation procedure relies on the \texttt{hyperopt} library~\cite{Bergstra_2015}, which systematically explores the (hyper)parameter selection space using Bayesian optimisation techniques~\cite{hyperopt}, evaluating model performance based on user-defined figures of merit to identify the optimal hyperparameters $\boldsymbol{\hat{\theta}}$, see Sect.~\ref{sec:hyperopt}.
As discussed there, the hyperoptimisation algorithm implemented in NNPDF4.0 is built on a $K$-fold cross-validation methodology, where all experimental data points entering the fit are divided into $n_{\rm K}$ folds.
For each hyperparameter configuration selected by \texttt{hyperopt}, also henceforth referred to as trial, $n_{\rm K}$ fits are performed, corresponding to the number of partitions. 
In each of these fits, the $p$-th fold is left out as a hold-out set, while the remaining folds are combined into a dataset, which is then split into training and validation sets as usual.
The final hyperoptimisation
loss function $L_{\rm hopt}$ (figure of merit) is then evaluated on the unseen hold-out set.

By means of the improvements in the NNPDF code described in this section and in App.~\ref{app:techimpr}, the NNPDF fitting methodology is now extended in three complementary directions, being able to:
\begin{enumerate}
    \item{Train hundreds of PDF replicas in parallel within each of the $n_{\rm K}$ fits.}
    \item{Run hyperparameter optimisations in parallel using multiple GPUs such that all trial runs share the same theory database.}
    \item{Utilise different hyperoptimisation metrics, with varying per-replica and per-fold statistics.}
\end{enumerate}

The availability of these three features within the NNPDF framework makes possible a combined multi-replica parallel hyperoptimisation algorithm with unprecedented performance and accuracy.
To support hyperoptimisation on multiple GPUs in parallel, we have adopted the distributed asynchronous hyperparameter scanning provided by \texttt{hyperopt}.
It uses a central MongoDB database to store hyperparameter choices and resulting (hyper-)losses for each trial. Multiple concurrent  processes, so-called \texttt{hyperopt} workers, execute trials sequentially and in between, communicate with the database to store hyperoptimisation metrics and load a new training configuration. 
In Fig.~\ref{fig:hyperopt_performance_improvement}, we explore the speedup of hyperoptimisation on a {\sc\small Snellius} GPU node with 4 accelerator cards.
These tests are conducted using the same settings as the final results to be shown in Sect.~\ref{sec:results}, with fits performed using either 50 or 100 PDF replicas. 

We observe that ``parallel" hyperopt runs with a single \texttt{hyperopt} worker are slightly more time-consuming than standard sequential runs, possibly because of communication overhead with the database. This can be significantly mitigated by allocating more workers, and thus allowing multiple trials to be executed simultaneously. Allocating more worker processes per GPU (light grey areas in Fig.~\ref{fig:hyperopt_performance_improvement}) does not seem to significantly impact the training speed, irrespective of the number of replicas being fitted. 
This indicates that the graphics cards are either fully utilised by a single training process or TensorFlow does not allow an efficient interleaving of our processing pipelines.
%

%
In this work we have not explored the possibility to extend the hyperoptimisation workload beyond 4 GPU's as this would require us to establish connections between nodes and communicate via HTTP, which is an unsupported procedure on our HPC cluster.
%
%

\begin{figure}[t]
    \centering
\includegraphics[width=0.49\linewidth]{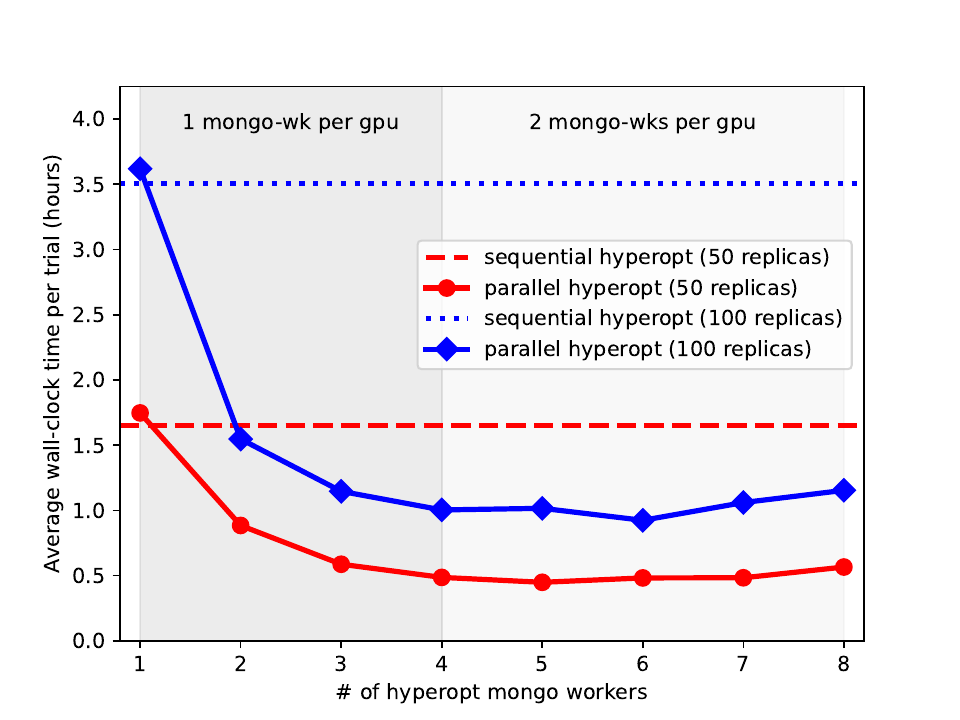}    
\includegraphics[width=0.49\linewidth]{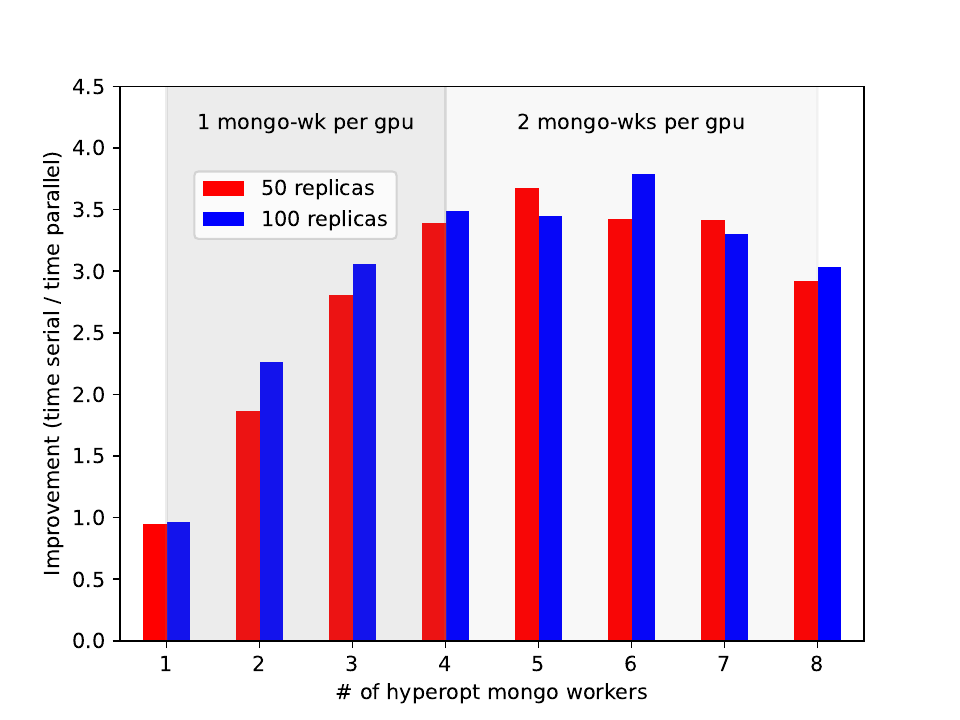}
    \caption{Comparison of the performance of the sequential and the parallel GPU-based hyper-optimisation runs. 
    For the parallel runs, we employ the distributed asynchronous hyperparameter scanning with MongoDB as implemented in \texttt{hyperopt} package.
    Left: The average wall-clock time per hyperopt trial as a function of the number of hyperopt mongo-worker processes, where each worker is responsible for completing one  trial,
    for both $N_{\rep}=50$ and 100 replicas.
    Right: The performance improvement in the parallel hyperoptimisation, quantified by the ratio between the wall-clock time per trial run run when running fits sequentially as opposed to running in parallel, also for $N_{\rep}=50$ and 100 replicas, and as a function of the number of hyperopt mongo-workers.
    }
\label{fig:hyperopt_performance_improvement}
\end{figure}

\section{Results}
\label{sec:results}

We now present the variant of the NNPDF4.0 global analysis based upon the new hyperparameter selection methodology described in Sect.~\ref{sec:hyperopt} and implemented according to the methodological improvements presented in Sect.~\ref{sec:implementation}.
First, we describe the theory settings
and experimental inputs that define this NNPDF4.0 variant.
Then, we detail the hyperparameter scan procedure and discuss how the models selection is performed. 
Finally, we present the resulting PDF fit, compare it with its
NNPDF4.0 baseline counterpart, and briefly assess the implications for LHC phenomenology.

\paragraph{Theory settings and experimental input.}
The NNPDF4.0 variant presented in this work is based on the NNPDF4.0 fits presented in~\cite{NNPDF:2021njg,NNPDF:2021uiq} and then extended in a series of recent papers to account for QED corrections~\cite{NNPDF:2024djq},  missing higher order uncertainties (MHOUs)~\cite{NNPDF:2024dpb}, and for theory calculations based on  approximate N$^3$LO (aN$^3$LO) QCD cross-sections~\cite{NNPDF:2024nan}.
These variants are based on an updated theory pipeline~\cite{Barontini:2023vmr} assembled around the {\sc\small EKO}~\cite{Candido:2022tld, candido_2022_6340153}, 
{\sc\small YADISM}~\cite{Candido:2024rkr, candido_2023_8066034}, and {\sc\small PineAPPL}~\cite{Carrazza:2020gss, christopher_schwan_2024_13951776} programs.
These are also the same settings used in other recent NNPDF studies such as the dedicated LO, NLO, and NNLO PDF sets 
for Monte Carlo event generators~\cite{Cruz-Martinez:2024cbz}.
In all cases, the {\sc\small YADISM} implementation of the FONLL general-mass variable-flavour-number scheme~\cite{Forte:2010ta} described in~\cite{Barontini:2024xgu} for heavy quarks is adopted.

In order to facilitate the direct comparison with the previous NNPDF hyperoptimisation strategy, here the fits will be carried at NNLO accuracy and do not contain neither QED effects, a photon PDF, MHOUs, nor aN$^3$LO corrections.
In the following, when we refer to NNPDF4.0, it should be understood as the fit variant described in App.~A of~\cite{NNPDF:2024djq} and based on the same hyperparameters $\boldsymbol{\hat{\theta}}$ determined in the original NNPDF4.0 study. 
The total dataset is composed by $n_{\rm dat}=4616$ points, of which more than 1000 cross-sections correspond to the LHC experiments, in particular
678, 473, and 116 for ATLAS, CMS, and LHCb datasets respectively.

\paragraph{Hyperparameter scan.}
The partitioning of the experimental datasets into folds follows the NNPDF4.0 choices, see Table~3.2 in~\cite{NNPDF:2021njg}.
These partitions are selected such that each fold is representative of the global dataset in terms of both kinematic coverage and of process types.
The hyperparameter space considered here is described in Table~\ref{tab:restricted_hyperparam_space}. 
As compared to the NNPDF4.0 study, here we consider a restricted space in which we hyperoptimise on the neural network architecture (for its two internal layers) and the optimiser hyperparameters (algorithms, norm clipping, and the value of the learning rate).
For the fixed hyperparameters, their values are set to be the same as the baseline 
configuration from NNPDF4.0.
We adopt $n_{\rm K} = 4$, and for each fold the PDF models were trained
using $N_{\rm rep}=100$ parallel replicas fitted simultaneously on GPUs.
As explained in Sect.~\ref{sec:hyperopt_nnpdf40}, for each  fits and set of hyperparameters we evaluate the values of $\chi^{2(k)}_{{\rm PDF},p}$, Eq.~(\ref{eq:chi2definition_6}), and of $\varphi^2_{\chi^2_p}$, Eq.~(\ref{eq:chi_estimator_pfold}), in the hold-out (non-fitted) datasets, and store them to evaluate the associated hyperoptimisation metrics, Eq.~(\ref{eq:hyperopt_2_replicas_PDF}) and Eq.~(\ref{eq:hyperoptloss_l3}) respectively. 
The hyperparameter scans presented below were conducted on single {\sc\small Snellius} GPU node and, following Sect.~\ref{sec:implementation}, employing 4 \textrm{hyperopt} workers, each allocated to a separate GPU.
%

\begin{table}[!t]
\centering
\small
\renewcommand{\arraystretch}{1.40}
\begin{tabularx}{\textwidth}{Xlll} 
\toprule
Parameter          & NNPDF4.0          & \multicolumn{2}{c}{This work, sampled range} \\ 
\cmidrule{3-4}
                   &                   &  min.                  & max. \\
\midrule
Internal neural network architecture & [$n_1$=25, $n_2$=20]   & $n_1,n_2$=10           & $n_1,n_2$=45           \\
optimiser                &  \textsc{Nadam}        & \textsc{Nadam}         & \textsc{Adam}          \\
Clipnorm                 & \(6.0 \times 10^{-6}\) & \(1.0 \times 10^{-7}\) & \(1.0 \times 10^{-4}\) \\
Learning rate            & \(2.6 \times 10^{-3}\) & \(1.0 \times 10^{-4}\) & \(1.0 \times 10^{-2}\) \\
\midrule
Maximum \# training epochs        & 17000                  &   \multicolumn{2}{c}{17000}                                              \\
Activation function      & \textsc{Tanh}          &               \multicolumn{2}{c}{\textsc{Tanh}}           \\
NN initializer              & \textsc{Glorot Normal}&               \multicolumn{2}{c}{\textsc{Glorot Normal}}                         \\
Stopping patience        & 0.1                    &  
 \multicolumn{2}{c}{0.1}\\
Initial positivity  multiplier      & 185                    &                \multicolumn{2}{c}{185}                                  \\
Initial integrability multiplier    & 10                     &             \multicolumn{2}{c}{10}                                     \\
\bottomrule
\end{tabularx}
\vspace{0.3cm}
\caption{The hyperparameter space considered in this work, compared with the corresponding ranges from NNPDF4.0 (see also Table~9 of~\cite{NNPDF:2021njg}).
We consider the internal neural network architecture, the optimiser, the value of the clipnorm parameter, and the learning rate.
As compared to NNPDF4.0, here a subset of the hyperparameters (lower part of the table) is kept fixed to their reference values and is not included in the scan.
}
\label{tab:restricted_hyperparam_space}
\end{table}

\paragraph{Hyperparameter selection.}
We have generated $n_{\rm trials} = 200$ sets of hyperparameter configurations $\boldsymbol{\hat{\theta}}$.
Out of these 200 sets of hyperparameters, we select the
best $n_{\rm hopt}=10$ ones,
following the combined selection strategy outlined in Sect.~\ref{sec:hyperopt_new}, namely:

\begin{enumerate}

\item First, we evaluate the metric $L_{\rm hopt}^{(\chi^2_{\rm pdf})}$ over the $n_{\rm trials} = 200$ trials and determine the $1\sigma$ range $\mathcal{R}$, measuring its fluctuations from the minimum defined in Eq.~(\ref{eq:step3_combined_algorithm}).

\item Then, for each of the acceptable models within the range $\mathcal{R}$, we evaluate the second hyperoptimisation metric $L_{\rm hopt}^{(\varphi^2)}$ and select the $n_{\rm hopt}=10$ trials with the lowest values.

\item For each of the $n_{\rm hopt}=10$ selected models, we generate equal samples of
$N_{\rm rep}/n_{\rm hopt}=20$ replica fits each, producing in total $N_{\rm rep} = 200$ replicas.

\item This fit with $N_{\rm rep} = 200$ replicas, built upon the combination of the best $n_{\rm hopt}=10$ sets of hyperparameters, constitutes the sought-for NNPDF4.0 variant and is denoted as {\tt nnpdf40\_newhyperopt} in the following.

\end{enumerate}

As discussed in Sect.~\ref{sec:hyperopt_new}, other options to select hyperparameters based on their values of $L_{\rm hopt}^{(\varphi^2)}$ are possible, such as the exponential weighting defined in Eq.~(\ref{eq:exponential_weighting}).
The investigation of such alternative model selection algorithms is left for future work.

The results of this procedure are displayed in Fig.~\ref{fig:loss-hyperopt}, with the values of $L_{\rm hopt}^{(\chi^2_{\rm pdf})}$ and $L_{\rm hopt}^{(\varphi^2)}$ for each trial shown in the left and right $y$-axes respectively for each of the hyperparameter $\boldsymbol{\hat{\theta}}$ considered.
The hyperparameters associated to the selected models, as well as the associated values of the hyperoptimisation metrics, are collected in Table~\ref{tab:best_models_hyperparams}. 
Trials leading to a poor fit quality, with $L_{\rm hopt}^{(\chi^2_{\rm pdf})}\ge 5$, are by default discarded; these add up to 20\% of the considered hyperparameter space. 

\begin{figure}[!t]
    \centering
\includegraphics[width=0.99\textwidth]{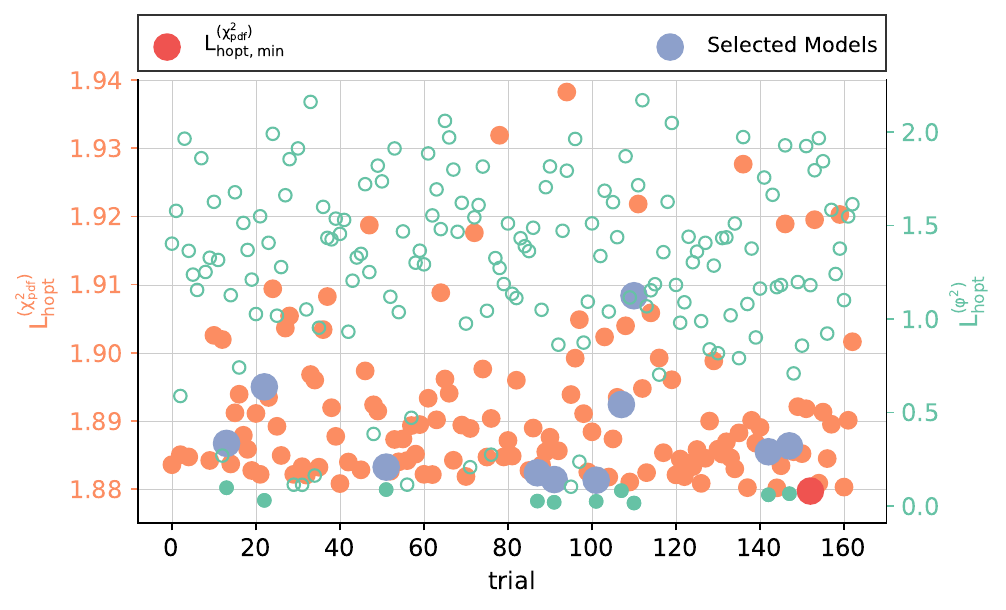}
    \caption{Graphical representation of the hyperoptimisation loss $L_{\rm hopt}^{(\chi^2_{\rm pdf})}$, Eq.~(\ref{eq:hyperopt_2_replicas_PDF}), as a function of the trials, where each trial corresponds to a different hyperparameter $\boldsymbol{\hat{\theta}}$ configuration, with orange filled circles in the left $y$-axis.
    We also show, for each trial, the value of the loss function $L_{\rm hopt}^{(\varphi^2)}$, Eq.~(\ref{eq:hyperoptloss_l3}), as empty green circles in the right $y$-axis. 
    The trial with the lowest value of $L_{\rm hopt}^{(\chi^2_{\rm pdf})}$ is encircled in red while the selected $n_{\rm hopt}=10$ trials are encircled in blue. These models are selected based on the trials with the lowest values
    of $L_{\rm hopt}^{(\varphi^2)}$ which are represented by the green filled circles.
    Due to the values of the hyperoptimisation losses being concentrated around 1.88 we only show a narrow range of $L_{\rm hopt}^{(\chi^2_{\rm pdf})}$.
    See Table~\ref{tab:best_models_hyperparams} for the hyperparameters associated to the selected models.
    }
    \label{fig:loss-hyperopt}
\end{figure}

\begin{table}[t]
	\scriptsize
	\centering
	\renewcommand{\arraystretch}{1.9}
	\begin{tabularx}{\textwidth}{X c c c c c c c c c c}
    \toprule
    Trial ID & 13 & 22 & 51 & 87 & 91 & 101 & 107 & 110 & 142 & 147 \\
    \midrule
    Architecture & [38, 35] & [28, 29] & [24, 22] & [42, 26] & [24, 23] & [25, 20] & [39, 22] & [43, 26] & [28, 18] & [44, 23] \\
    optimiser & Adam & Nadam & Nadam & Nadam & Nadam & Nadam & Nadam & Nadam & Nadam & Nadam \\
    Clipnorm [$10^{-5}$] & 5.48 & 4.75 & 0.60 & 2.12 & 0.70 & 4.40 & 3.73 & 0.72 & 5.21 & 2.42 \\
    Learning~Rate [$10^{-3}$] & 1.95 & 4.10 & 6.52 & 5.43 & 3.08 & 3.97 & 3.31 & 2.08 & 9.25 & 5.84 \\
    \midrule
    $L_{\rm hopt}^{(\chi^2_{\rm pdf})}$ & 1.89 & 1.88 & 1.88 & 1.88 & 1.88 & 1.88 & 1.88 & 1.88 & 1.89 & 1.88 \\
   $L_{\rm hopt}^{(\varphi^2)}$ [$10^{-1}$] & 5.87 & 1.13 & 9.53 & 9.31 & 1.13 & 0.25 & 0.19 & 0.24 & 8.57 & 9.22 \\
    \bottomrule
    \end{tabularx}
	\vspace{0.3cm}
	\caption{
	The hyperparameters corresponding to each of the 10 selected models
   from Fig.~\ref{fig:loss-hyperopt} with the corresponding values of the hyperoptimisation metrics.
    The trial ID is arbitrary. 
	}
\label{tab:best_models_hyperparams}
\end{table}

A distinctive feature that arises from  Fig.~\ref{fig:loss-hyperopt} is that a large number of models lead to an equally satisfactory performance as measured by the two hyperoptimisation metrics.
This is to be expected: given the large size of the hyperparameter space being considered, there should exist a large population of points that lead to quasi-equivalent performance at the fit level. 
These considerations further justify the relevance of carrying out the NNPDF fits in terms of a wide population of different hyperparameters $\boldsymbol{\hat{\theta}}$, rather than in terms of just on a single set as done in the current methodology. 

From Fig.~\ref{fig:loss-hyperopt} one also observes the potential limitations of a hyperoptimisation strategy based exclusively on first-moment metrics: among trials that provide a satisfactory description of the central values of the non-fitted data, one finds large variations in the associated PDF uncertainties for the same set of observables.
This implies that one may select trials which generalise in a non-optimal manner, and in particular which may underestimate PDF uncertainties in the non-fitted datasets. 
The combined strategy deployed in this work instead achieves the best of both scenarios: given trials that describe equally well the data in the non-fitted folds, the algorithm select those with the largest PDF uncertainties.

\paragraph{Fit quality and statistical estimators.}
The {\tt nnpdf40\_newhyperopt} fit is composed by $N_{\rm rep}=200$ replicas which combine the selected best $n_{\rm hopt}=10$ sets of hyperparameters.
This fit variant can then be directly compared with the NNPDF4.0 baseline, namely the NNLO QCD-only fit described in App.~A of~\cite{NNPDF:2024djq}. 
The only difference between the two fits is hence the hyperparameters $\boldsymbol{\hat{\theta}}$ defining the models.

Table~\ref{tab:tab-chi2} provides an overview of the $\chi^2$ values (per data point), both total and classified into process type, for the {\tt nnpdf40\_newhyperopt} fit compared with the corresponding NNPDF4.0 baseline. 
We find that the overall fit quality is somewhat improved as compared to the baseline, and similar effects are observed at the level of individual groups of processes. 
For the global dataset with $n_{\rm dat}=4618$, the total $\chi^2$ decreases a bit, from 1.16 to 1.15, in the fit variant based on the new hyperoptimisation procure. 
This improvement is spread among all process types, without any single cross-section type dominating the $\chi^2$ decrease. 

\begin{table}[!tb]
  \footnotesize
  \centering
  \renewcommand{\arraystretch}{1.5}
  \begin{tabularx}{\textwidth}{Xrrr}
  \toprule
  \multirow{1}{*}{Dataset}
  & \multicolumn{1}{c}{$n_{\rm dat}$}
  & \multicolumn{1}{c}{NNPDF4.0}
  & \multicolumn{1}{c}{{\tt nnpdf40\_newhyperopt}}
  \\
  \midrule
  DIS NC (fixed-target)
  & 
  973
  &
  1.26
  &
  1.25
  \\
  DIS CC (fixed-target)
  & 
  907
  &
  0.86
  &
  0.86
  \\
  DIS NC (collider)
  & 
  930
  &
  1.19
  &
  1.18
  \\
  DIS CC (collider)
  & 
  81
  &
  1.28
  &
  1.27
  \\
  Drell-Yan (fixed-target)
  &
  195
  &
  1.00
  &
  1.02
  \\
  Tevatron $W,Z$ production (inclusive) 
  & 
  65
  &
  1.09
  &
  0.99
  \\ 
  LHC $W,Z$ production (inclusive) 
  &
  483
  &
  1.37
  &
  1.35
  \\
  LHC $W,Z$ production ($p_T$ and jets)
  & 
  150
  &
  0.98
  &
  0.98
  \\
  LHC top-quark pair production
  & 
  66
  &
  1.21
  & 
  1.24
  \\
  LHC jet production
  & 
  500
  &
  1.26
  & 
  1.24
  \\
  LHC isolated $\gamma$ production
  & 
  53
  &
  0.77
  &
  0.76 
  \\
  LHC single $t$ production
  & 
  17
  & 
  0.36
  &
  0.37
  \\
  \midrule
  Total
  & 
  4618
  & 
  1.16
  &
  1.15
  \\
\bottomrule
\end{tabularx}
    \vspace{0.2cm}
  \caption{The $\chi^2$ values by process type in {\tt nnpdf40\_newhyperopt} 
  compared to the NNPDF4.0 baseline.
  }
  \label{tab:tab-chi2}
\end{table}

Table~\ref{tab:tab-estimators} presents the same comparison as in Table~\ref{tab:tab-chi2} now at the level of other statistical estimators relevant for the interpretation 
of the PDF fit results.
In particular, we show the average over replicas of the training and validation loss functions, $\la E_{\rm tr}\ra_{\rm rep}$ and $\la E_{\rm val}\ra_{\rm rep}$ respectively, the average value of the training length $\la {\rm TL}\ra_{\rm rep}$, the average value over replicas of the total figure of merit $\la \chi^2_{\rm tot}\ra_{\rm rep}$, and the value of the $\varphi^2_{\chi^2}$ estimator defined in Eq.~(\ref{eq:chi_estimator}) and measuring the PDF uncertainties in the fitted cross-sections in units of the experimental data uncertainties.

\begin{table}[!tb]
  \footnotesize
  \centering
  \renewcommand{\arraystretch}{1.7}
\begin{tabularx}{\textwidth}{Xrr}
  \toprule
  \multirow{1}{*}{Estimator}
  & \multicolumn{1}{c}{NNPDF4.0}
  & \multicolumn{1}{c}{{\tt nnpdf40\_newhyperopt}}
  \\
  \midrule
  $\la E_{\rm tr}\ra_{\rm rep}$
  & 
  $2.27\pm 0.05$
  &
  $2.25 \pm 0.05$	
  \\
  $\la E_{\rm val}\ra_{\rm rep}$
  & 
  $2.37\pm 0.10$
  &
  $2.37\pm 0.10$	
  \\
  $\la {\rm TL}\ra_{\rm rep}$
  & 
  $12200\pm 2200$
  &
  $10300 \pm 2000$	
  \\
  $\la \chi^2_{\rm tot}\ra_{\rm rep}$
  & 
  $1.194\pm 0.014$
  &
  $1.187 \pm 0.015$	
  \\
   $\varphi^2_{\chi^2}$
  & 
  $0.154\pm 0.004$
  &
  $0.175 \pm 0.002$	
  \\
\bottomrule
\end{tabularx}
    \vspace{0.2cm}
  \caption{Same as Table~\ref{tab:tab-chi2} for other relevant statistical estimators assessing fit quality.
  }
  \label{tab:tab-estimators}
\end{table}

From the comparisons in Table~\ref{tab:tab-estimators}, one finds that {\tt nnpdf40\_newhyperopt} is, for most estimators, statistically equivalent to the baseline fit.
One noticeable difference appears to be at the level of the $\varphi^2_{\chi^2}$ estimator, which increases its value by around 10\% in the former case.
Given its interpretation in terms of the PDF uncertainties in the fitted cross-sections, recall Eq.~(\ref{eq:varphi_estimator_2}), one may conclude that the new hyperparameter optimisation algorithm leads to a increase of the PDF uncertainties in the data region by around 5\% on average as compared to the NNPDF4.0 baseline.
This moderate increase, despite the radically different approach to select model hyperparameters, further confirms the robustness of the PDF uncertainty estimate obtained in the original NNPDF4.0 analysis.
As shown below, a similar qualitative trend, more marked in overall magnitude, is observed in the extrapolation regions of the PDFs.

\paragraph{Impact on PDFs.}
Having compared {\tt nnpdf40\_newhyperopt} with NNPDF4.0 at the level of statistical estimators, we now compare them at the level of the PDFs and their uncertainties.
Fig.~\ref{fig:pdf-comparisons} displays a comparison of the PDFs in the baseline NNPDF4.0 global analysis and in  {\tt nnpdf40\_newhyperopt}  at $Q=100$ GeV,  normalised to the central value of the former.
 One observes an excellent consistency between the two fits, in agreement with the findings of the statistical estimators in Tables~\ref{tab:tab-chi2} and~\ref{tab:tab-estimators}.
 The shift in the central values of the new fit is in all cases much smaller than the PDF uncertainties of NNPDF4.0.
 For the light quark and antiquark PDFs, the new hyperoptimisation procedure leads to somewhat larger PDF uncertainties both in the data region (as expected from the increase in $\varphi^2_{\chi^2}$ in Table~\ref{tab:tab-estimators}) as well as in the small-$x$ extrapolation region, where few experimental constraints are available.
 Interestingly, the charm and gluon PDFs are left essentially unaltered in the fit based on the new hyperopt. 

\begin{figure}[!t]
	\centering
    \includegraphics[width=0.495\textwidth]{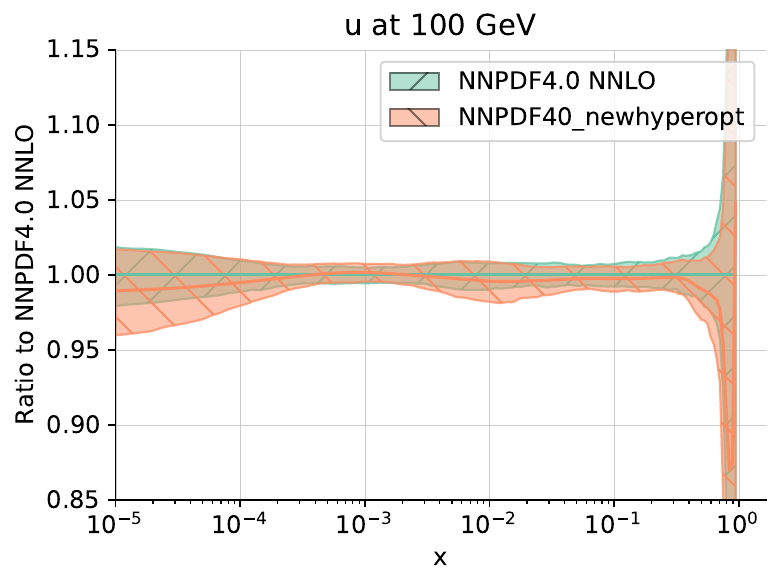}
    \includegraphics[width=0.495\textwidth]{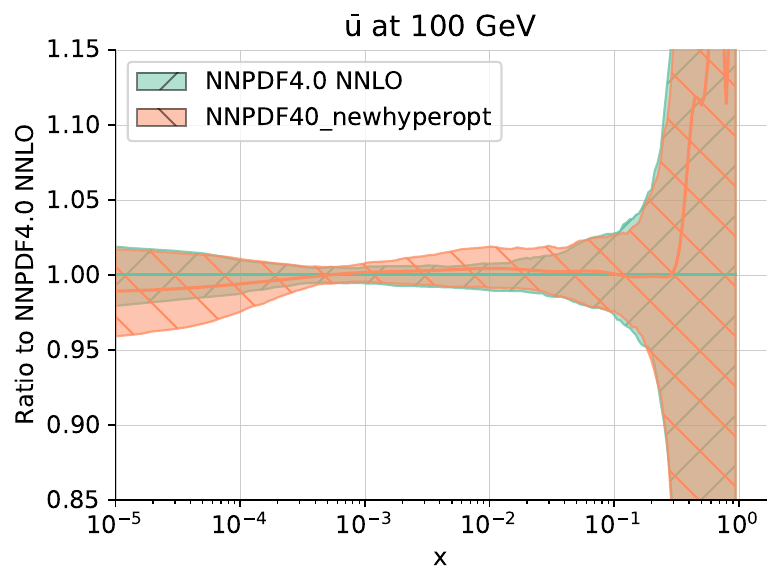}
    \includegraphics[width=0.495\textwidth]{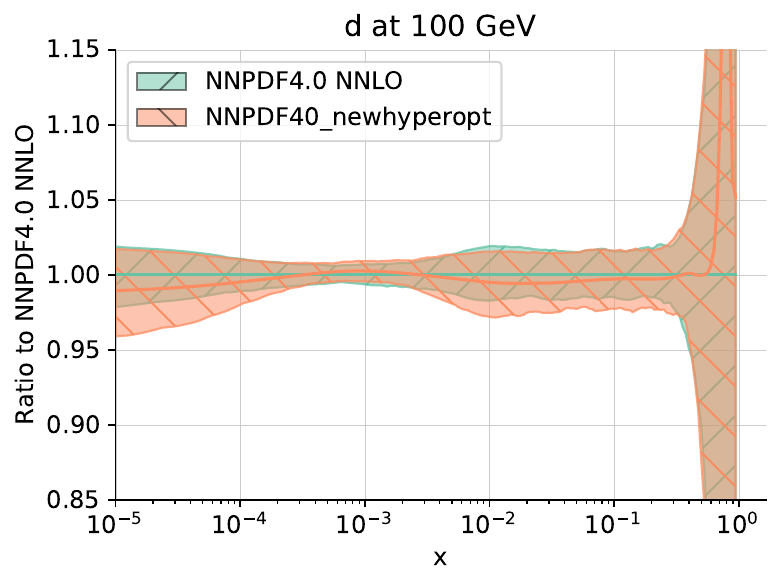}
    \includegraphics[width=0.495\textwidth]{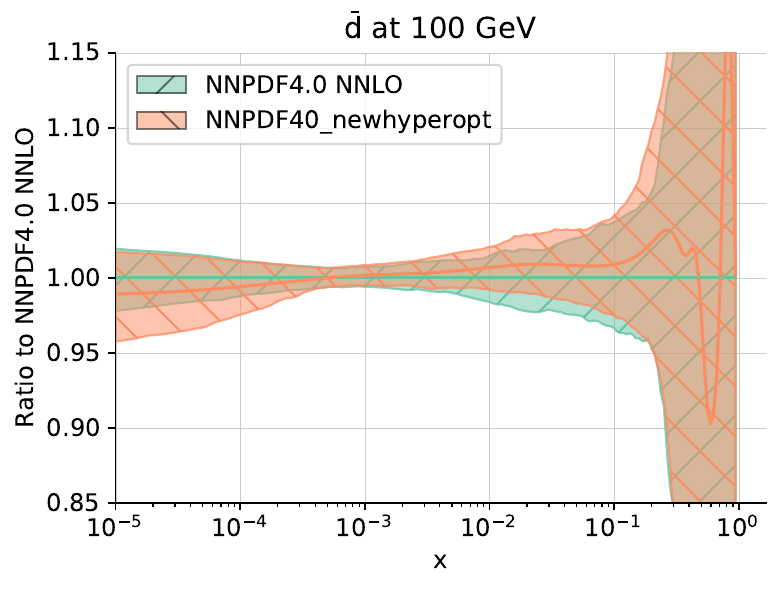}
    \includegraphics[width=0.495\textwidth]{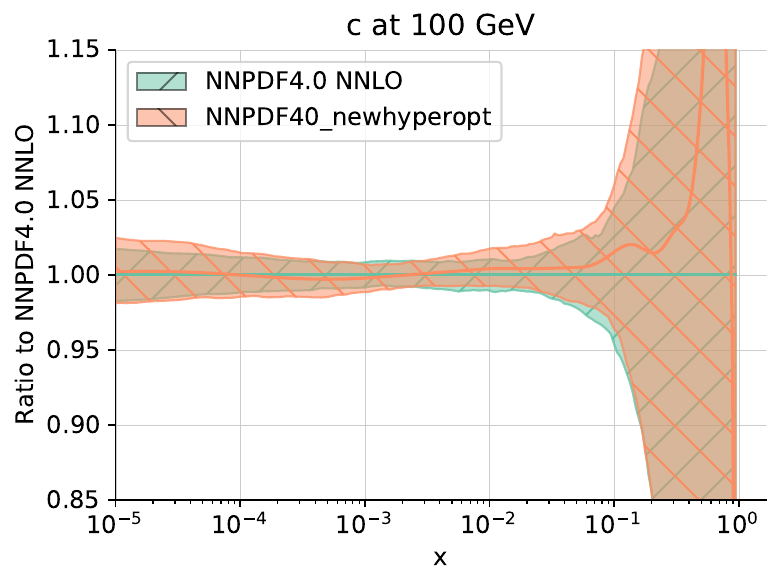}
    \includegraphics[width=0.495\textwidth]{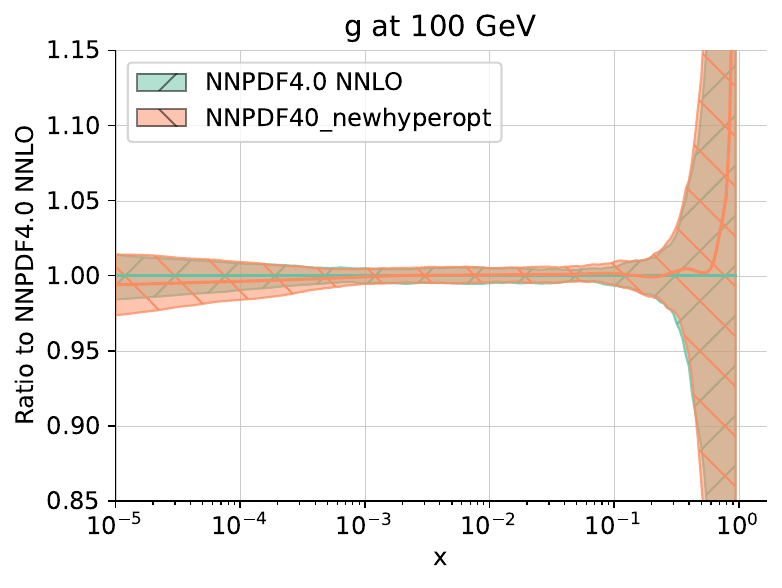}
	\caption{
	Comparisons of the anti-up, up, anti-down, down, charm, and gluon PDFs between NNPDF4.0 and {\tt nnpdf40\_newhyperopt}.
    Results are presented at $Q=100$ GeV as a function of $x$, the 
 uncertainty bands represent the $68\%$ confidence interval, and are shown normalised to the central value of NNPDF4.0.
	}
	\label{fig:pdf-comparisons}
\end{figure}

Next, a similar comparison as in Fig.~\ref{fig:pdf-comparisons} but now at the level of the absolute PDF uncertainties is shown in Fig.~\ref{fig:uncertainty-comparisons}.
As mentioned above, no changes are found for the gluon and charm quark PDFs.
For the light quark and antiquark PDFs, the {\tt nnpdf40\_newhyperopt} fit leads to a moderate increase of the PDF uncertainties in the data region and a larger one in the small-$x$ extrapolation region with $x\lsim 10^{-3}$, where there are currently limited constraints on light quark flavour separation.
No sizeable differences between the PDF uncertainties in the two fits in the large-$x$ region are observed. 

\begin{figure}[!t]
	\centering	\includegraphics[width=0.495\textwidth]{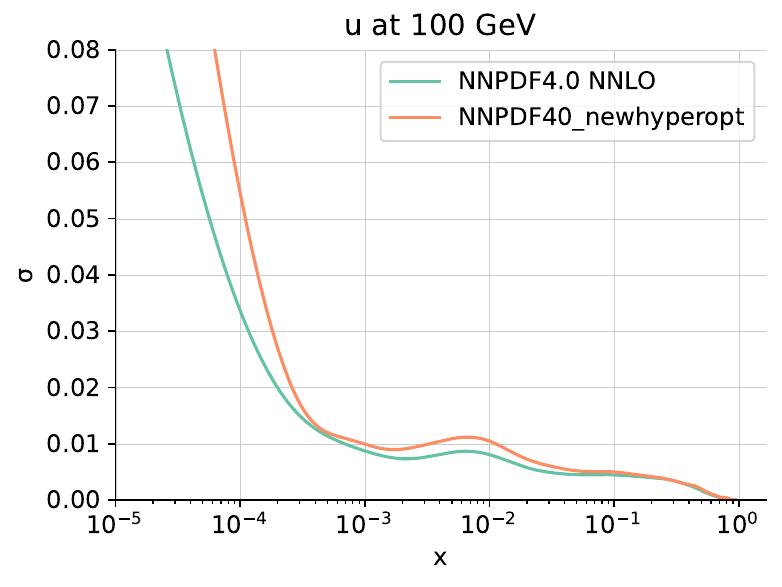}
\includegraphics[width=0.495\textwidth]{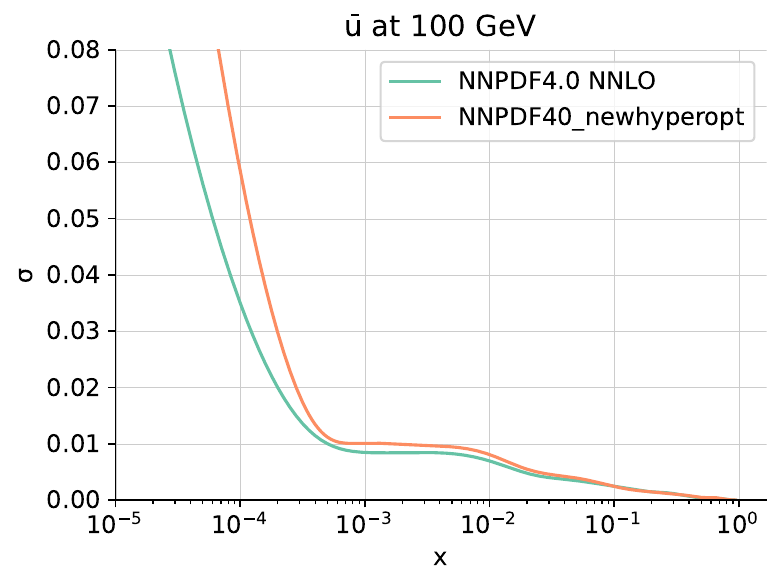}
\includegraphics[width=0.495\textwidth]{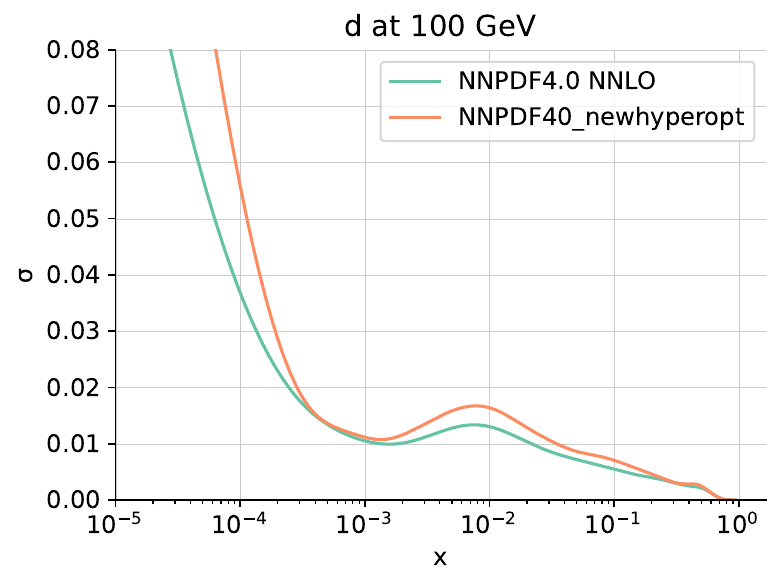}
\includegraphics[width=0.495\textwidth]{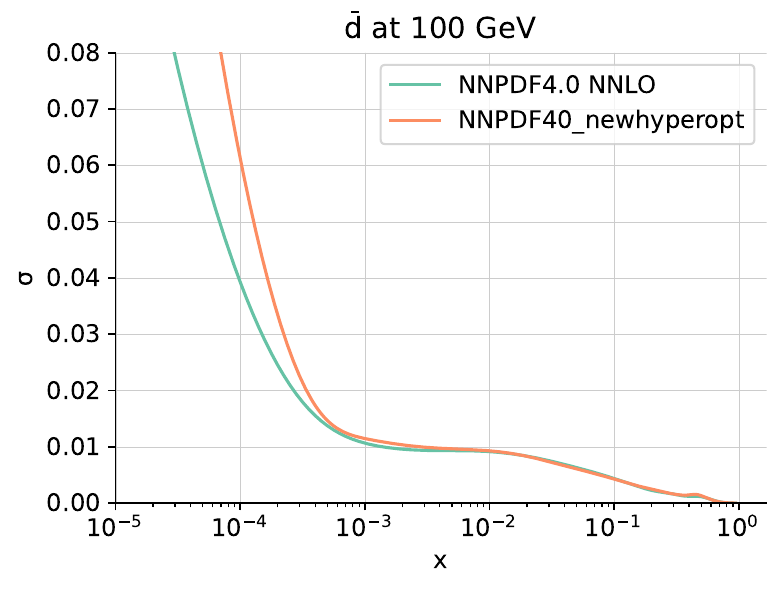}
\includegraphics[width=0.495\textwidth]{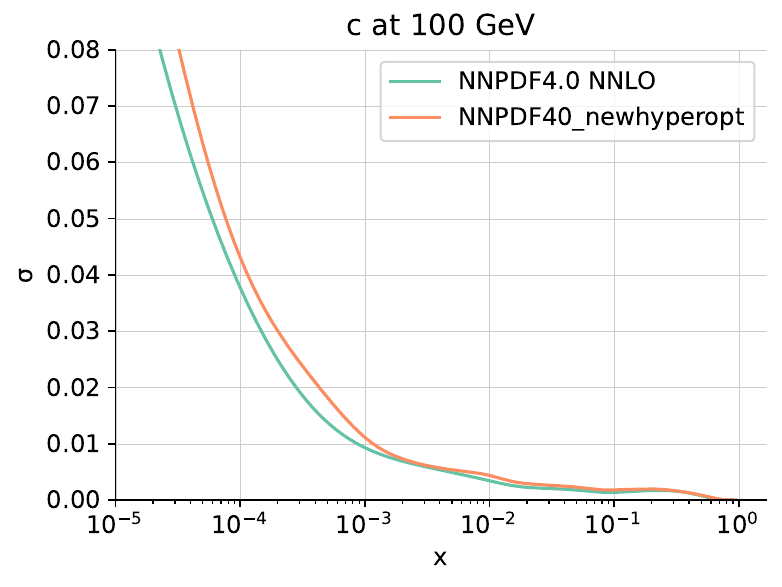}
\includegraphics[width=0.495\textwidth]{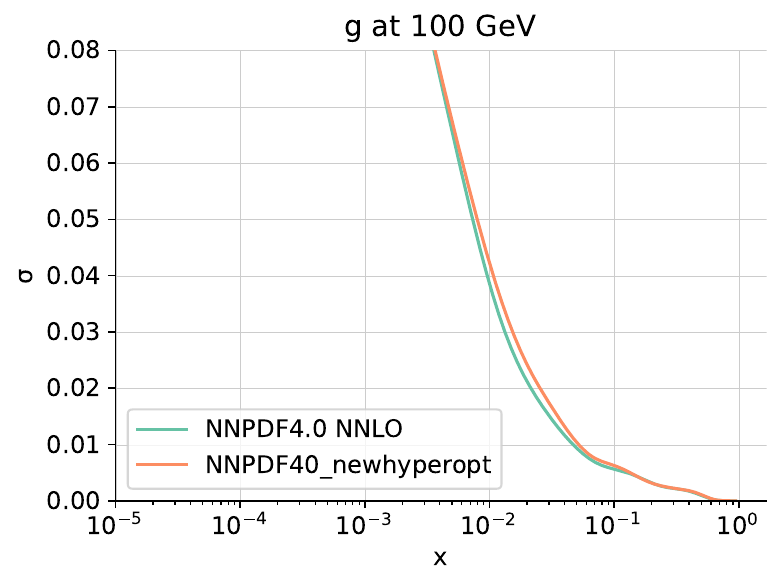}
\caption{Same as Fig.~\ref{fig:pdf-comparisons} now for the one-sigma absolute PDF uncertainties.
}
\label{fig:uncertainty-comparisons}
\end{figure}

One concludes that the hyperparameter optimisation algorithm developed in this work leads to results which are consistent with the original NNPDF4.0 implementation, while displaying somewhat larger uncertainties specially in small-$x$ extrapolation region.

\paragraph{Implications for LHC predictions.}
Finally let us briefly comment on the implications of our analysis for theoretical predictions of LHC cross-sections.
From Fig.~\ref{fig:pdf-comparisons} one can expect that the effects of the new hyperoptimisation algorithm will modify the LHC cross-sections computed with NNPDF4.0  via a moderate increase of the PDF uncertainties in
the small- and intermediate-mass regions.
To assess this effect in a qualitative manner, we show in Fig.~\ref{fig:luminosity-comparisons} a comparison between the partonic luminosities at the LHC computed with NNPDF4.0 and with  {\tt nnpdf40\_newhyperopt}.
Results are evaluated at  $\sqrt{s} = 14~\mathrm{TeV}$, differential in the invariant mass $m_X$ of the produced final state and integrated over its rapidity $y_X$.

\begin{figure}[!t]
	\centering
    \includegraphics[width=0.495\textwidth]{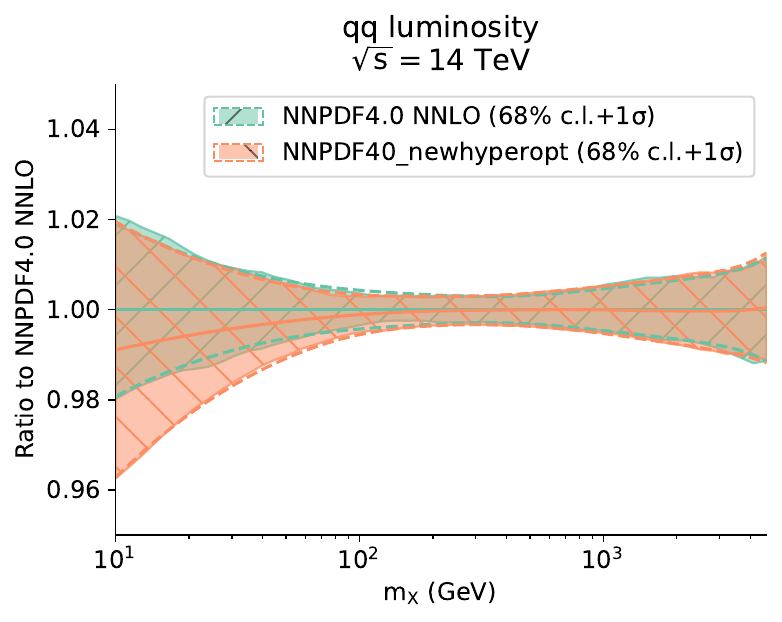}
    \includegraphics[width=0.495\textwidth]{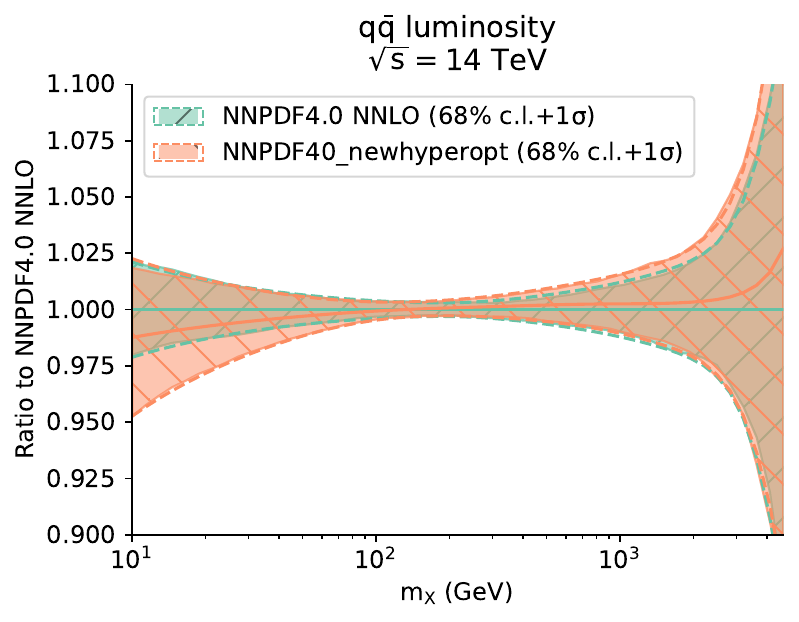}
    \includegraphics[width=0.495\textwidth]{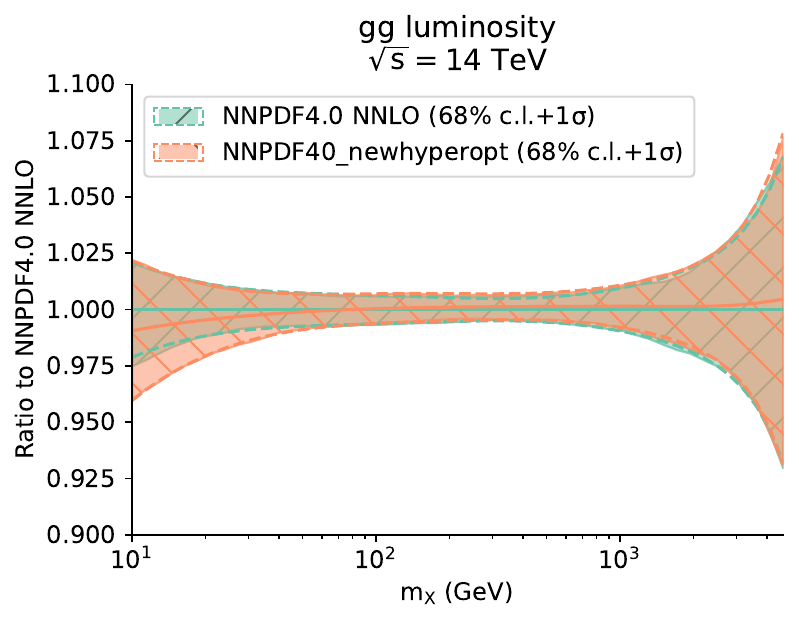}
    \includegraphics[width=0.495\textwidth]{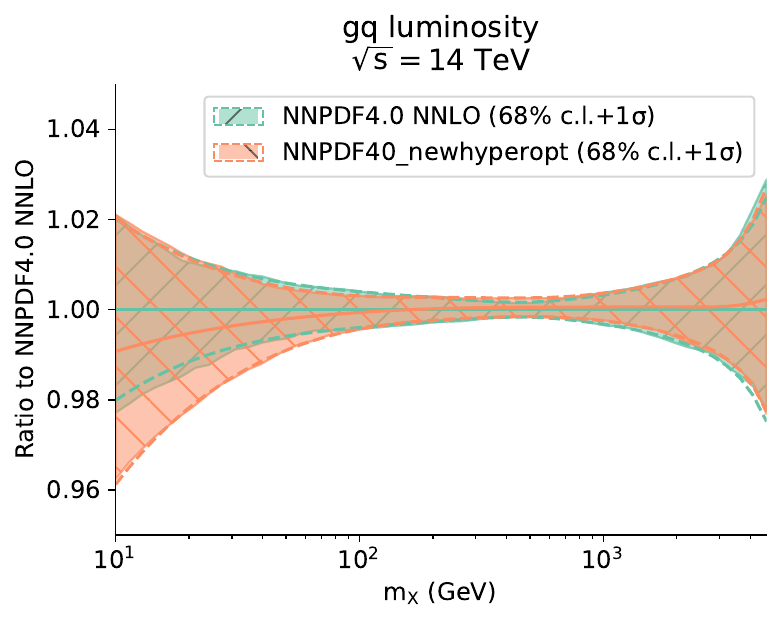}
	\caption{Same as Fig.~\ref{fig:pdf-comparisons} for the partonic luminosities at the LHC with $\sqrt{s} = 14~\mathrm{TeV}$, differential in the invariant mass $m_X$ of the produced final state and integrated over its rapidity $y_X$.
 We display the quark-quark, quark-antiquark, gluon-gluon, and quark-gluon luminosity combinations. 
}
\label{fig:luminosity-comparisons}
\end{figure}

Consistently with the associated PDF comparisons, excellent agreement is found at the level of the partonic luminosities both in terms of central values and of PDF uncertainties.
The main differences are the increase of PDF errors in the small-$m_X$ region, corresponding to the small-$x$ extrapolation region.
Within the data region, a small increase of the PDF uncertainties in the quark-antiquark luminosity, at the 10\% level, is the most noticeable effect.
We conclude that theoretical predictions for LHC cross-sections evaluated with NNPDF4.0 are robust upon the implementation of the improved  hyperparameter optimisation algorithm.
\section{Summary and outlook}
\label{sec:summary}

In this work we have presented a novel approach to the automated determination of hyperparameters in the context of the training of machine learning models, and applied it to carry out an improved extraction of the quark and gluon substructure of the proton within the NNPDF4.0 framework.
Our strategy is based on defining target metrics for hyperoptimisation constructed in terms of the complete probability distribution on model space, and not only on its first moment as done in previous studies.
Our findings further establish the robustness of the NNPDF methodology with respect to variations in the ML model prior assumptions, and motivate their adoption as baseline for future NNPDF releases. 

Our approach makes it possible to select PDF models with the best generalization power, defined as those models providing a satisfactory description of out-of-sample data while at the same time maximizing the PDF uncertainties when extrapolated to non-fitted observables. 
Another important difference as compared to NNPDF4.0 is that now the PDF fit is constructed from a distribution of different model hyperparameters, rather than selecting a single model and then training $N_{\rm rep}$ replicas with identical hyperparameters. 

Realising this new approach to hyperparameter optimisation has required a very significant restructuring and technical performance optimisation of the NNPDF framework, in particular enabling the parallel running of hundreds of PDF replicas and aligning the codebase for its execution both in CPUs and in GPUs. 
As a result, a speed-up of up to two orders of magnitude is achieved, which, irrespective of applications to hyperparameter optimisation, will be beneficial for NNPDF studies that involve a very large number of fits, such as the determination of the strong coupling $\alpha_s(m_Z)$ based on the correlated replica method~\cite{Ball:2018iqk}, level-2 closure test studies~\cite{NNPDF:2014otw}, and eventual extractions of the PDFs together with SM parameters such as the top quark mass or the $W$ boson mass at the LHC and beyond. 
Given the increasing availability of GPU compute clusters, this new functionality will  accelerate future studies of proton substructure and LHC phenomenology. 
Complementing this speed-up, the same technical improvements enable large reductions in memory requirements, energy consumption, and overall costs of the NNPDF model training.

Beyond unpolarized PDFs, our new hyperparameter optimisation strategy will also be applicable in the context of polarized PDF fits within the NNPDF framework~\cite{Nocera:2014gqa}, and it has already been successfully deployed for the upcoming NNPDFpol2.0 determination~\cite{Hekhorn:2024jrj}.
Likewise, an integrated QCD analysis combining polarized and unpolarized PDFs with nuclear PDFs~\cite{AbdulKhalek:2022fyi} and with fragmentation functions~\cite{Bertone:2017tyb} would be characterized by a much more complex hyperparameter space as compared to the individual fits carried out so far, and hence will benefit from the progress presented here.
Such an  integrated QCD fit~\cite{Khalek:2021ulf} represents a crucial requirement to fully exploit the physics program of the upcoming Electron Ion Collider~\cite{AbdulKhalek:2021gbh} to start taking data in the early 2030s.

Beyond applications to proton structure, given that this new strategy to hyperparameter selection is fully general, we expect it to be of interest for other ML practitioners in different areas of physics and beyond. 
Indeed, as ML applications grow in prominence within the modern fundamental science landscape, novel strategies to ascertain the robustness of the model training results and to reduce the need for human intervention (and hence possible bias) in ML model selection become more important than ever.
The insights gained in this work concerning the adaptation of the NNPDF codebase to parallel model training with GPUs may also be of relevance for other ML researchers facing a similar problematic, and who also would like to benefit from the increasing availability of hardware accelerators.

\vspace{0.5cm}
\begin{center}
\noindent\rule{0.7\linewidth}{0.3mm}
\end{center}
\vspace{0.5cm}

\noindent
The results of this work have been implemented in the code  available in the public NNPDF repository:
\begin{center}
\url{https://github.com/NNPDF/nnpdf}
\end{center}
starting from v4.0.9, as well as in the alpha releases of 4.1, see also the online code documentation:
\begin{center}
\url{https://docs.nnpdf.science/}
\end{center}

\subsection*{Acknowledgements}

We are grateful to our colleagues of the NNPDF Collaboration for multiple fruitful discussions about hyperparameter optimisation in ML.
The work of J.~R. is partially supported
by NWO, the Dutch Research Council. 
The work of T.~R.~R. and J.~R. is partially supported by an Accelerating Scientific Discoveries (ASDI2021) grant from the Netherlands eScience Center (NLeSC), grant number 027.020.G05.
J.~R. is grateful to the Theory Division of CERN for hospitality during the final part of this project.
R.~S. is supported by the U.K. Science and Technology
Facility Council (STFC) consolidated grants ST/T000600/1 and ST/X000494/1.
Computing resources used in this work were provided by the Dutch national e-infrastructure with the support of the SURF Cooperative using grant noa. EINF-5725, EINF-9109 and EINF-11006.
\appendix 
\section{Technical improvements in the NNPDF framework}
\label{app:techimpr}

Here we provide further details on the technical improvements implemented in the NNPDF framework.
Given that the NNPDF code is based on {\tt Keras}, and through it can access both {\tt TensorFlow} and {\tt PyTorch}, it was already possible to run the fitting code seamlessly on GPUs.
However, initial benchmarks found that the resulting performance and memory use of the multi-replica fit was much worse than expected.
Indeed, despite GPU occupancy and memory bandwidth being well below their theoretical capacity, increasing the number of replicas resulted in longer overall training times as compared to sequential fits.
To improve this situation, it was necessary to carry out a number of performance optimization as explained in Sect.~\ref{sec:implementation}, resulting in GPUs being able to simultaneously accommodate hundreds of network replicas, leading to a significant speed-up of the fits without the memory consumption exploding. 
The resulting performance enhancements for multi-replica PDF fits running on GPUs and the decrease in peak memory consumption have been summarised in Fig.~\ref{fig:multirepbenchmark}.
Further to the code improvements already described in Sect.~\ref{sec:implementation}, the following optimisation efforts have also been implemented:

\paragraph{Optimised FastKernel tables:}
Every time the PDF parameters are modified, the hadronic cross-sections need to be reevaluated by convoluting the PDFs with the DGLAP evolution kernels and with the partonic matrix elements.
Evaluating these cross-sections involves multi-dimensional integrals that are very computational intensive, and hence it is not feasible to repeat this process on the fly during a PDF fit which involves a very large number of iterations.
To bypass this limitation, NNPDF is based on the so-called FK-tables~\cite{Bertone:2016lga}, by which these convolutions are interpolated in terms of a tensor contraction with pre-computed kernels encoding the contributions of both DGLAP evolution and of the hard-scattering matrix elements,  as indicated by Eqns.~(\ref{eq:predict})--(\ref{eq:predict_dis}).

A global NNPDF fit requires a number of FK-tables of the order of the number of datasets $N_{\rm dsets}$ being considered.
Since by default the optimal interpolation grid was determined separately for each dataset, the PDFs (and hence the underlying neural networks) had to be evaluated at $N_x^{\rm (tot)} \sim N_{\rm dsets}\times N_x$ values, with $N_x=50$ being the number of $x$-grid nodes in an individual FK-table, resulting in  $N_x^{\rm (tot)} \sim 4000$ for a typical PDF fit with $N_{\rm dsets}\sim 80$ datasets fitted.
Such configuration shifts the computational cost of the fit from CPU-hours to memory requirements, necessitating about an hour and 3 GB of RAM for a typical fit.

To reduce the memory footprint and improve the computational efficiency, we have updated the NNPDF fitting framework to make use of FK tables based on {\sc \small PineAPPL}~\cite{Carrazza:2020gss, christopher_schwan_2024_13951776}.
Among other advantages, this new format ensures a common $x$-grid for the FK-tables irrespective of the dataset, such that in each iteration of the fit the neural networks only need to be evaluated $N_x=50$ (rather than $N_x^{\rm (tot)} \sim 4000$) times. 

\paragraph{Model stacking:}
In most machine learning applications parallelising over GPU cores is advantageous due to the depth and width of the networks involved, which makes the evaluation of the neural network output the computational bottleneck.
Within a NNPDF fit, instead, the parallelisation has to occur over the ensemble of replicas, since the most expensive part of the calculation is not the evaluation of the network output itself, but rather that of the hadronic cross-sections through the convolution of the PDFs with the FK-tables, Eqns.~(\ref{eq:predict})--(\ref{eq:predict_dis}).
To achieve this, we stack the replicas together before the convolution with the FK-tables and then separate them 
immediately afterwards, Fig.~\ref{fig:stacked_replicas}.
This feature allows the NNPDF code to exploit the GPU kernels of \texttt{TensorFlow} and the provided primitives for tensor algebra, while separately applying optimization replica per replica.

\paragraph{Memory management:}
Due to the different memory layout of the multi-replica fits, the original alignment of the tensors entering the FK-table convolution of Eq.~(\ref{eq:predict}) used in the NNPDF framework was sub-optimal, as it resulted in tensor contractions along non-contiguous indices preventing efficient re-utilisation of cached memory. 
To improve memory management, we now make the replica index a row-like index when more than one replica is requested and evaluate tensor contractions from left to right.
In addition, for hadron collider observables involving a double convolution, every FK-table needs to be contracted with the partonic luminosity tensor generated by a convolution of both incoming (and typically set to be the same) PDFs.
This causes a problem in multi-replica fits, since the luminosity tensors of each replica are independent of each other, giving raise to high memory usage.
To prevent this, when running with more than one replica simultaneously, since the FK-table is common for all replicas, it is more efficient to decompose beforehand the FK-table channel axis into two axes to be contracted separately with each PDF.
This feature ensures that a potentially large increase in memory consumption occurring in the multi-replica fits is prevented. 

\paragraph{Data masking:}
In previous incarnations of the FK-table formalism, for instance those used in NNPDF4.0, it was  advantageous to mask the FK-table eliminating entries not required to evaluate the physical cross-sections (for instance, excluded by kinematic cuts), significantly reducing their size.
Such a masking procedure had however the undesirable effect of generating objects separated in memory. 
To prevent this issue, with the new generation of {\sc \small PineAPPL}-based FK-tables, 
this mask is only applied when computing the loss as indicated in Eq.~(\ref{eq:chi2definition_masking}).
This modification results in an increased memory consumption in single-replica fits, nevertheless this increment is more than offset by the optimised FK-tables described above.
It is necessary, nevertheless, to make a compromise regarding datasets with a single datapoint, such as total cross-sections, which for fits based on parallel replicas are always included in the training set.

\paragraph{Replica-dependent seeds:}
In compliance with the open science and FAIR paradigms, all results presented in a scientific publication, in the case at hand PDF determinations, should be independently reproducible by external parties. 
In order to ensure that the NNPDF framework can be run in different machines and configurations resulting in the exact sets of PDFs as that presented in our publications, the initialization of the fitting procedure (that is, setting the initial values of the weights and thresholds, for example) is always performed as if the PDFs were to be run sequentially.
This choice incurs in a small penalty in efficiency, but guarantees that independent runs of the code are completely reproducible provided single datapoint datasets are not considered (see above). 
Even when single datapoint datasets are considered, the effect is typically much smaller than any change in the code initialization seeds.

\bibliographystyle{utphys}
\bibliography{new_hyperopt}

\end{document}